\documentclass[11pt]{article}

\usepackage[final]{acl}

\usepackage{times}
\usepackage{latexsym}

\usepackage[T1]{fontenc}

\usepackage[utf8]{inputenc}

\usepackage{microtype}

\usepackage{inconsolata}

\usepackage{graphicx}

\usepackage{amsmath}
\usepackage{subcaption}
\usepackage{booktabs}
\usepackage{amssymb}
\usepackage{placeins}
\usepackage{xurl}  %
\usepackage[T1]{fontenc}

\title{EviRerank: Adaptive Evidence Construction for Long-Document LLM Reranking}

\author{
  \textbf{Minghan Li}\textsuperscript{1}\thanks{Corresponding author: \texttt{mhli@suda.edu.cn}},
  \textbf{Eric Gaussier}\textsuperscript{2},
  \textbf{Juntao Li}\textsuperscript{1},
  \textbf{Guodong Zhou}\textsuperscript{1}
\\
  \textsuperscript{1}School of Computer Science and Technology, Soochow University, Suzhou, China
\\
  \textsuperscript{2}Univ. Grenoble Alpes, CNRS, Grenoble INP, LIG, 38000 Grenoble, France
\\
  \small{
  \texttt{\{mhli,ljt,gdzhou\}@suda.edu.cn}
  \quad
  \texttt{Eric.Gaussier@univ-grenoble-alpes.fr}
  }
}

\begin{document}
\maketitle

\begin{abstract}

Decoder-only LLM rerankers struggle with long documents: inference is costly and relevance signals can be diluted by irrelevant context. Motivated by an attention analysis indicating a consistent degradation trend when non-relevant text is appended, we propose \textbf{EviRerank}, an evidence-based long-document reranking framework for decoder-only LLMs. EviRerank (i) scores document blocks with a lightweight selector (BM25, bi-encoder, or cross-encoder), (ii) constructs a compact reranking context under a hard token cap by dynamically budgeting evidence blocks with \textbf{Adaptive Evidence Budgeting (AEB)} and adding a global summary cue via \textbf{Summary Augmentation (SA)}, and (iii) reranks with a decoder-only LLM. Across TREC DL'19, DL'22, DL'23, and MLDR-zh, EviRerank consistently outperforms full-document LLM reranking and strong block-selection baselines while substantially reducing the required input length. On TREC DL'19, EviRerank reaches up to \textbf{0.744} nDCG@10 while substantially reducing reranking latency. Additional RankZephyr-7B experiments confirm transfer to \textbf{listwise reranking} under matched context budgets.

\end{abstract}

\section{Introduction}
\label{sec:intro}

Large language models (LLMs)~\cite{touvron2023llama} have become strong rerankers (e.g., RankLLaMA~\cite{ma2024fine}) for web search and other retrieval settings, thanks to advances in modeling capacity and generalization. However, real-world documents are often long, and decoder-style LLMs incur rapidly growing inference costs as input length increases. Moreover, when relevant evidence is sparse and scattered, the reranker input may fail to cover the decisive signals, leading to unstable ranking quality.

\begin{figure}[t]
  \centering
  \includegraphics[width=\linewidth]{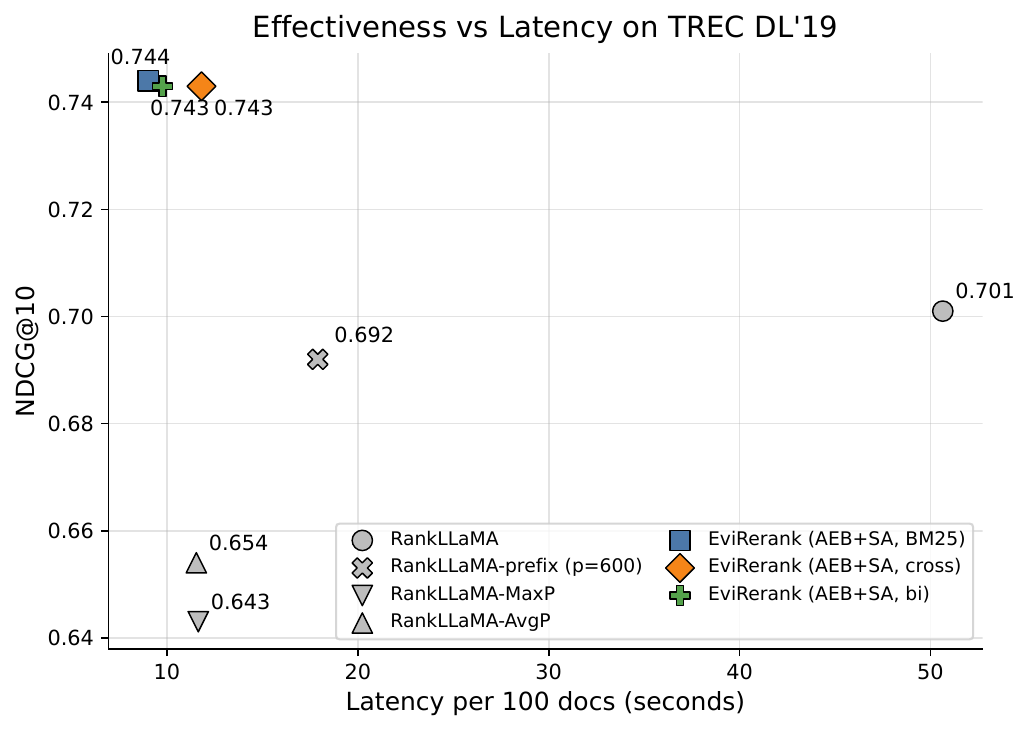}
  \caption{Effectiveness-efficiency trade-off on DL'19. We plot nDCG@10 against reranking latency per 100 documents (top-left better).}
  \label{fig:effecspeed}
  \vspace{-6pt}
\end{figure}

A common workaround is to decompose a document into smaller blocks and rerank based on a subset \cite{li2023power, li2022bert}. This ``block-first'' strategy introduces a new question: \emph{how should we construct an evidence context under a hard token cap?} Most existing block-based pipelines (e.g., KeyB \cite{li2023power} / IDCM-style \cite{DBLP:conf/sigir/HofstatterMZCH21} selection) primarily address \emph{which} blocks to keep, and often rely on an implicit fixed allocation rule (e.g., always taking a fixed number of top blocks or always filling the budget). However, documents differ greatly in information density: some contain one or two highly relevant blocks followed by many weak ones, while others have many moderately relevant blocks. As a result, a fixed allocation wastes tokens on low-utility tail blocks for the former, yet may still fail to allocate enough evidence for the latter.

In this work, we propose \textbf{EviRerank}, a lightweight evidence-guided reranking framework for decoder-only LLMs that constructs a compact reranker input under a hard maximum token cap. Given a query-document pair, EviRerank first scores document blocks with a local scorer (e.g., BM25 \cite{robertson2009probabilistic}, bi-encoder, or cross-encoder), and then performs evidence construction by allocating the capped context budget to block evidence and global cues. Crucially, we introduce {Adaptive Evidence Budgeting (AEB)}, a density-aware budgeting rule that determines {how much} block evidence to include: instead of always filling the cap, it halts further budget allocation when additional blocks provide low marginal utility, yielding a query-adaptive evidence length. In addition, we incorporate {Summary Augmentation (SA)} within the same cap to inject global document cues, and isolate its effect with controlled ablations. Experiments on TREC DL'19, DL'22, DL'23, and MLDR-zh show consistent gains and a better accuracy--efficiency trade-off. Additional RankZephyr-7B experiments show that the same evidence-construction layer transfers to listwise LLM reranking under matched context budgets. We focus on document reranking quality, rather than downstream answer generation.

\paragraph{Research questions.}
We study five questions: \textbf{RQ1 Mechanism:} What does attention reveal about long-document evidence construction? (Appendix~\ref{sec:attentionMap}) \textbf{RQ2 Generalization:} Do gains hold across datasets and across pointwise and listwise LLM reranking? \textbf{RQ3 Selectors:} How do BM25/bi-/cross-encoder selectors affect the effectiveness-efficiency trade-off? \textbf{RQ4 SA:} Does Summary Augmentation improve effectiveness under the same cap? \textbf{RQ5 Efficiency:} What are the latency and training-memory benefits?

\paragraph{Contributions.}
Our main contributions are:
\begin{itemize}
\item \textbf{Evidence construction for long-document LLM reranking.} We cast long-document reranking as constructing a compact evidence context under a token cap before LLM scoring. EviRerank is compatible with different block selectors and can be used as a construction layer for both pointwise and listwise LLM rerankers.
\item \textbf{Adaptive Evidence Budgeting (AEB).} We introduce a model-agnostic, normalized-score ratio stopping rule that adaptively determines how much block evidence to include.
\item \textbf{Summary Augmentation (SA).} We add a compact summary cue within the same cap and evaluate it with controlled baselines to rule out ``just more tokens'' effects.
\item \textbf{Results and analysis.} We show consistent gains on DL'19, DL'22, DL'23, and MLDR-zh, validate transfer to RankZephyr-7B listwise reranking, and provide controlled ablations and efficiency analyses.
\end{itemize}

The remainder of this paper is organized as follows. Section~\ref{sec:related} reviews related work. Section~\ref{sec:method} presents EviRerank, including AEB and SA. Section~\ref{sec:experimentSetting} and \ref{sec:experimentResults} describe the experimental settings and main results, followed by ablations and efficiency analyses in Section~\ref{sec:ablation}.

\section{Related Work}
\label{sec:related}

\subsection{Neural IR Selectors: Cross- and Bi-encoders}
Neural IR with PLMs is commonly instantiated as (i) cross-encoders that jointly encode query-text pairs for accurate reranking~\cite{devlin-etal-2019-bert,nogueira2019passage}, (ii) bi-encoders that learn dense representations for efficient scoring or retrieval~\cite{karpukhin2020dense,reimers-gurevych-2019-sentence}, and (iii) late-interaction models that balance token-level matching and efficiency~\cite{khattab2020colbert,santhanam2022colbertv2}. In this work, we leverage these paradigms as lightweight local selectors (BM25/bi/cross) to score document blocks before LLM reranking.

\subsection{Long-Document Reranking}
Long documents challenge Transformer rerankers due to quadratic attention and diluted relevance signals. Efficient/sparse Transformers extend context length (e.g., Longformer~\cite{beltagy2020longformer}, BigBird~\cite{zaheer2021big}, and related sparse designs~\cite{child2019}), and IR-specific variants such as TKL~\cite{hofstatter2020local} and QDS-Transformer~\cite{jiang2020long} further tailor attention patterns. A complementary line segments documents into passages and aggregates signals, using simple pooling~\cite{dai2019deeper} or learned aggregation (e.g., PARADE~\cite{li2023parade}). More recently, block selection reduces noise by identifying salient passages before applying stronger rankers: IDCM~\cite{DBLP:conf/sigir/HofstatterMZCH21}, ICLI~\cite{li2022bert}, and KeyB~\cite{li2023power} report strong results by selecting and concatenating blocks.

Different from prior work that primarily treats block selection as a pre-processing step for BERT-style rerankers or simple concatenation, we study \emph{budget-aware evidence construction} tailored to decoder-only LLM rerankers: selecting query-salient evidence blocks, composing them into a compact context under a fixed token budget, and optionally augmenting it with a lightweight summary cue.

\subsection{LLM-based Reranking}
Finetuned LLM rerankers such as RankLLaMA~\cite{ma2024fine} demonstrate strong effectiveness, but long-document reranking remains challenging due to context limits and high inference cost. Listwise LLM rerankers such as RankGPT-style prompting~\cite{sun2023chatgpt} and RankZephyr-7B~\cite{pradeep2023rankzephyr} introduce a related budget tension because a single context window is shared across multiple candidate documents. Recent work also explores progressive adaptation~\cite{zhang2024two} and context compression for efficiency (e.g., PERank~\cite{liu2024leveraging}). In contrast to full-document LLM reranking or aggregation-only pooling, our approach explicitly constructs a compact evidence context before LLM scoring. We further evaluate whether this evidence-construction layer transfers from RankLLaMA-style pointwise reranking to RankZephyr-style listwise reranking.

\section{EviRerank: Evidence-Guided LLM Reranking with Adaptive Budgeting}
\label{sec:method}

\begin{figure}[t]
\centering \includegraphics[width=0.98\linewidth]{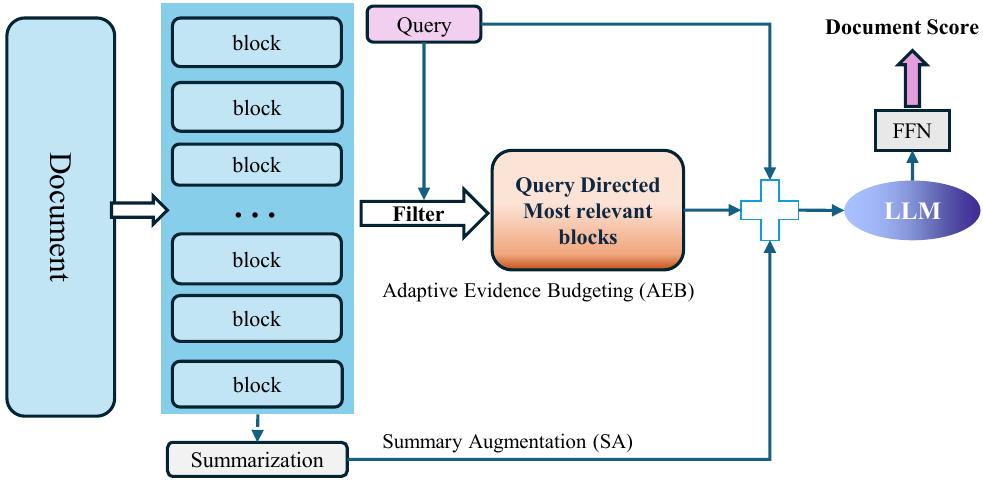}
  \caption{Architecture of \textbf{EviRerank}. The framework composes the reranker input from query-focused evidence blocks, optionally augmented with a compact query-agnostic summary augmentation (SA). Adaptive Evidence Budgeting (AEB) enables early stopping under the same budget.}
  \label{fig.architecture}
  \vspace{-5pt}
\end{figure}

Fig.~\ref{fig.architecture} illustrates \textbf{EviRerank}, an evidence-guided reranking framework for long documents. EviRerank decomposes each document into blocks, scores blocks with a lightweight local scorer, and then composes a compact LLM input for a decoder reranker. Two components are optional and can be toggled independently: (i) \textbf{Adaptive Evidence Budgeting (AEB)} for information-density-aware early stopping, and (ii) \textbf{Summary Augmentation (SA)} that injects a short, query-agnostic summary cue under the same total budget.

\subsection{Passage Segmentation}
We segment each document into blocks using the CogLTX decomposition method \citep{ding2020cogltx}, which favors strong punctuation boundaries and uses dynamic programming to ensure each block length is bounded. Following \citet{li2023power}, we set the maximum block length to $B{=}63$. For consistency with the final reranker, block segmentation uses the reranker tokenizer (Llama2/Llama3 accordingly), and we extend the punctuation set for Chinese to better detect sentence boundaries.

\subsection{Local Block Scoring}
\label{sec:keyb2-local}
After splitting a long document into blocks, EviRerank scores each block locally and then performs global reranking with a decoder LLM. The local scoring stage is lightweight (orders of magnitude cheaper than LLM inference) and \emph{model-agnostic}: any standard IR scorer can be plugged in. In our experiments, we instantiate three widely used options; full formulas and details are provided in Appendix~\ref{app:local-ranking}.

\noindent\textbf{BM25 (term matching).} We adopt the standard BM25 with $(k_{1},b)$ and smoothed IDF (computed with \texttt{scikit-learn} \cite{pedregosa2011scikit}). BM25 provides a strong term-matching baseline and is robust for long documents. For Chinese texts we apply standard word segmentation before scoring.

\noindent\textbf{Cross-encoder (interaction).} A pretrained encoder (e.g., BERT~\cite{devlin-etal-2019-bert}) takes the concatenated \texttt{[query; block]} as input; the \texttt{[CLS]} representation is fed to a linear head to yield a relevance score. Because each block is short, per-block inference is still efficient while retaining fine-grained token-level interactions.

\noindent\textbf{Bi-encoder (dense retrieval).} A shared encoder maps queries and blocks to vectors; local relevance is computed by a dot product (or cosine). Block embeddings can be precomputed offline, making scoring efficient at runtime while capturing semantic similarity beyond lexical overlap.

\subsection{Composing the LLM Input: Evidence Blocks with a Summary Cue}
\label{sec:compose-llm-input}

\paragraph{Setup.}
Given a query $Q$ and a document segmented into blocks $\{b_i\}_{i=1}^{n}$, a local scorer (BM25/bi-encoder/cross-encoder; Sec.~\ref{sec:keyb2-local}) produces block relevance scores $s_i$ and block lengths $L(b_i)$. Let the decoder reranker token cap for document-side content be $B$. We allocate this cap between evidence blocks and the summary cue: $B = B_K + B_S$, where $B_K$ is the maximum budget for evidence blocks $K$, and $B_S$ is the maximum budget for the summary cue $S$ (with $B_S=0$ when SA is disabled). Let $k_s$ be the number of blocks used to form the summary cue.

\paragraph{Step 1: Adaptive Evidence Budgeting (AEB) via information-density-aware early stopping.}
\label{para:dynamic_density}
Naive packing fills the token budget by appending blocks, but long documents often exhibit highly skewed \emph{information density} (a few decisive blocks followed by a low-utility tail). To avoid diluting strong evidence with weak blocks, we introduce \textbf{AEB}, which \emph{early-stops} evidence packing when the marginal relevance of newly considered blocks falls below a threshold, producing a query-adaptive evidence length that does not necessarily fill the cap $B_K$.

\paragraph{Score normalization.}
Raw block scores $s_i$ can have heterogeneous scales across local scorers (e.g., BM25 vs.\ cross-encoder logits), making ratio-based stopping brittle. We transform scores into a bounded, monotonic scale $\tilde{s}_i$ before early stopping:
\begin{equation}
\label{eq:keyb2_score_norm}
\tilde{s}_i = g(s_i),
\end{equation}
where $g(\cdot)$ is either identity (\textsc{None}) or Min-Max normalization:
\begin{equation}
\label{eq:keyb2_minmax}
g(s_i)=
\begin{cases}
s_i, & \textsc{None},\\[4pt]
\dfrac{s_i - s_{\min}}{s_{\max}-s_{\min}+\epsilon}, & \textsc{Min-Max},
\end{cases}
\end{equation}
with $s_{\min}=\min_j s_j$, $s_{\max}=\max_j s_j$, and a small constant $\epsilon$ (set to $10^{-12}$).

\paragraph{Dynamic selection with ratio-based stopping.}
Let $\pi$ sort blocks by descending normalized score $\tilde{s}$: $\tilde{s}_{\pi_1}\ge \cdots \ge \tilde{s}_{\pi_n}$. We scan blocks in this order and greedily add $b_{\pi_j}$ into $K$ if it fits the evidence budget ($T+L(b_{\pi_j})\le B_K$); otherwise, we stop (i.e., blocks are packed atomically rather than split).

To avoid selecting low-density tail blocks, we apply a ratio-based early stopping rule after selecting at least $m$ blocks:
\begin{equation}
\label{eq:keyb2_dynamic_stop}
\text{stop if } |K|\ge m \;\wedge\; \tilde{s}_{\pi_j}< \rho\,\tilde{s}_{\pi_1}.
\end{equation}
Here $\tilde{s}_{\pi_1}$ is the best normalized block score and $\tilde{s}_{\pi_j}$ is the current block score in the sorted order. The hyperparameter $\rho\in[0,1]$ controls how aggressively we truncate the long tail: once the score drops below a fixed fraction of the best block, we stop adding further blocks even if budget remains. Setting $\rho=0$ disables ratio-based stopping, reducing the procedure to budget-only selection.

Finally, we restore the original document order of selected blocks in $K$. As a safety measure, we apply a hard truncation after composing the final input (Step~3) to ensure the total document-side content does not exceed $B$ tokens, accounting for tokenization and prefix tokens.

\paragraph{Step 2: Summary Augmentation (SA) as a query-agnostic cue.}
We build a compact summary cue from blocks. We obtain block embeddings $\{e_i\}_{i=1}^{n}$ from the bi-encoder (shared with the selector by default) and cache them offline for efficiency, then compute the centroid
\begin{equation}
  \label{eq:keyb2_centroid}
  \hat{c} \;=\; \frac{\sum_i e_i}{\left\lVert \sum_i e_i \right\rVert},
\end{equation}
score each block by its similarity to the centroid
\begin{equation}
  \label{eq:keyb2_sum_score}
  s^{\text{sum}}_i \;=\; e_i \!\cdot\! \hat{c},
\end{equation}
and select the top-$k_s$ blocks by $s^{\text{sum}}_i$ (preserving original order) to form a short summary $S$.

\paragraph{Step 3: Compose and score with the decoder reranker.}
We first compose the document-side content as
\begin{equation}
  \label{eq:keyb2_Dtilde}
  \tilde{D} \;=\; K \;\Vert\; S ,
\end{equation}
and then enforce the hard token cap by truncation:
\begin{equation}
  \label{eq:keyb2_Dprime}
  D' \;=\; \mathrm{Trunc}_{B}\!\big(\tilde{D}\big),
\end{equation}
where $K$ is constructed under the evidence budget $B_K$ and $S$ under the summary budget $B_S$, with $B = B_K + B_S$.

\paragraph{Pointwise reranking.}
For the RankLLaMA-style pointwise instantiation used in our main experiments, we format the decoder-only reranker input as
\begin{equation}
  \label{eq:keyb2_input_format}
  \texttt{input} =
  \text{``query: \{Q\} document: \{D'\}''}.
\end{equation}
The final relevance score is produced from the last hidden state:
\begin{equation}
  \label{eq:keyb2_rsv}
  \mathrm{RSV}(Q,D) \;=\;
  \mathrm{Linear}\!\big(\mathrm{Decoder}(\texttt{input})[-1]\big).
\end{equation}

\paragraph{Listwise use.}
The evidence construction in Eqs.~\ref{eq:keyb2_Dtilde}--\ref{eq:keyb2_Dprime} is independent of the downstream scoring interface. For a candidate set $\{D_i\}_{i=1}^{N}$, we apply the same construction independently to each document to obtain compact representations $\{D'_i\}_{i=1}^{N}$. A listwise reranker can then consume these representations in place of full-document truncations. In Section~\ref{sec:rankzephyr}, we evaluate this setting with RankZephyr-7B while keeping its native listwise prompt and sliding-window reranking procedure unchanged.

\noindent\textbf{Rationale.} AEB adaptively allocates the evidence budget per query-document pair, avoiding low-utility tail blocks. SA adds a compact query-agnostic global cue that complements the selected evidence blocks. Both are lightweight: scoring operates on short blocks, and SA is a top-$k$ selection over cached block embeddings.

\section{Experimental Settings}
\label{sec:experimentSetting}

We evaluate \textbf{EviRerank} on multiple long-document reranking benchmarks and compare it with strong baselines, mainly targeting \textbf{RQ2}-\textbf{RQ5} introduced in Section~\ref{sec:intro}.

\subsection{Datasets}
We use four datasets emphasizing long or full-article documents:
\begin{itemize}
\item \textbf{TREC DL 2019} (document reranking) \cite{craswell2020overview}: MS~MARCO v1-based document reranking with human relevance judgments.
\item \textbf{TREC DL 2022} (document reranking): MS MARCO v2-based document reranking, used as an additional held-out validation benchmark.
\item \textbf{TREC DL 2023} (document reranking): built on MS~MARCO v2 with web-page style documents.
\item \textbf{MLDR-zh} \cite{chen2024m3}: the Chinese subset of MLDR, sourced from Wikipedia and Wudao \cite{yuan2021wudaocorpora}.
\end{itemize}
Dataset statistics are reported in Appendix~\ref{app:dataSetStat}.

\subsection{Baselines}
We compare against competitive first-stage and reranking systems. Baseline sets align with prior work while remaining consistent across datasets where applicable. Most baselines are described in Appendix~\ref{sec:moreBaselines}.

\paragraph{Controlled RankLLaMA baselines.}
On DL'19, we additionally include a matched-budget RankLLaMA-prefix control, which uses the same pointwise reranker but truncates each document to its first $p_{\max}=600$ document-side tokens.

\paragraph{RankLLaMA-MaxP / AvgP}
To align with block-based designs, we segment each document into blocks, compute per-block scores with RankLLaMA, and then aggregate: MaxP takes the maximum block score and AvgP averages block scores. These serve as strong aggregation-only counterparts to evidence selection and composition.

\paragraph{RankZephyr-7B.}
To evaluate whether EviRerank transfers beyond pointwise RankLLaMA-style scoring, we additionally evaluate RankZephyr-7B~\cite{pradeep2023rankzephyr}, a listwise LLM reranker. Within each document budget, full-document truncation and EviRerank use the same candidate set, context limit, prompting protocol, listwise window/stride, and decoding settings. Detailed settings are reported in Table~\ref{tab:rankzephyr-appendix}(a) of Appendix~\ref{app:rankzephyr-protocol}.

\subsection{Experimental Design and Implementation Details}
\label{sec:exper_design}

\paragraph{Variants of EviRerank.}
We evaluate \textbf{EviRerank} as an LLM-augmented reranker with fine-tuning. To ensure fair comparison with RankLLaMA (LLaMA2-7B\footnote{\url{https://huggingface.co/meta-llama/Llama-2-7b-hf}} backbone), we fine-tune EviRerank using the same backbone. EviRerank adopts local block scorers as modular components:
\[
\text{EviRerank}_{\text{BM25}},\quad \text{EviRerank}_{\text{bi}},\quad \text{EviRerank}_{\text{cross}},
\]
corresponding to BM25, bi-encoder, and cross-encoder selectors. We by default enable \textbf{Adaptive Evidence Budgeting (AEB)} (dynamic stopping) and \textbf{Summary Augmentation (SA)}.

\paragraph{Implementation summary.}
We use PyTorch~2.4 with CUDA~11.8 on a single NVIDIA A100 (40GB). We fine-tune with LoRA \cite{hu2021lora} while freezing the backbone, using pairwise hinge loss \cite{li2023power} and AdamW with warmup-decay in FP16. Default LoRA hyperparameters are $r{=}32$, $\alpha{=}64$, learning rate $5{\times}10^{-5}$, batch size 2, and gradient accumulation steps 8. All models train for one epoch on the same triplets. RankLLaMA uses gradient checkpointing (for longer inputs) and batch size 1; otherwise hyperparameters mirror EviRerank. All models are evaluated as rerankers over the same first-stage candidates (official runs for DL tracks; for MLDR-zh, BM25 top-$k$).

Training data and evaluation metrics are in Appendix~\ref{app:trainData}.

\paragraph{RankZephyr evaluation protocol.}
For the listwise evaluation, we use the \texttt{castorini/rank\_zephyr\_7b\_v1\_full} checkpoint with the official RankLLM prompt format and rerank the official top-100 candidates under a 4096-token context window. For EviRerank, each candidate document is independently converted to its compact evidence representation using the cross-encoder selector with AEB+SA before listwise reranking. The exact per-document budgets and window/stride settings are reported in Table~\ref{tab:rankzephyr-appendix}(a) of Appendix~\ref{app:rankzephyr-protocol}.

\paragraph{Input length and packing.}
Queries are truncated to 32 tokens following \cite{ma2024fine}. Unless otherwise specified, we enforce a hard maximum cap of $p_{\max}{=}600$ document-side tokens. Under this cap, we compose the reranker context from (i) query-focused evidence blocks and (ii) a lightweight summary cue (\textbf{SA}), while never exceeding the cap. By default, we allocate up to 480 tokens to evidence blocks and up to 120 tokens to the summary cue (within the same $p_{\max}$ cap). With \textbf{AEB}, evidence blocks are allocated dynamically: selection may terminate early when additional blocks provide low marginal utility, yielding a shorter query-adaptive evidence length under the same hard cap. For controlled ablations, we remove \textbf{SA} and reallocate its token budget to evidence blocks, while keeping the same $p_{\max}{=}600$ cap; we also compare against a 3-block random summary cue under the same cap (Table~\ref{tab:abl-summary}). In contrast, the full-document RankLLaMA baseline consumes up to 4,096 document-side tokens. The DL'19 matched-budget RankLLaMA-prefix control instead truncates each document to the same $p_{\max}=600$ document-side cap.

\paragraph{Block selectors and pretrained encoders.}
We use language-specific  cross-encoders and a multilingual bi-encoder as selectors in EviRerank (Table~\ref{tab:cross-bi-encoders}); Unless otherwise specified, the same bi-encoder is used for both block selection and summary cue construction.

\begin{table}[htbp]
  \centering
  \scriptsize
  \setlength{\tabcolsep}{2pt}
  \renewcommand{\arraystretch}{1.0}
  \begin{tabular}{@{}l l p{0.48\columnwidth}@{}}
  \toprule
  \textbf{Encoder} & \textbf{Lang} & \textbf{HF checkpoint} \\
  \midrule
  Cross & EN &
  \href{https://huggingface.co/cross-encoder/ms-marco-MiniLM-L-6-v2}
       {\texttt{cross-encoder/ms-marco-MiniLM-L-6-v2}} \\
  Cross & ZH &
  \href{https://huggingface.co/BAAI/bge-reranker-base}
       {\texttt{BAAI/bge-reranker-base}} \\
  Bi & EN/ZH &
  \href{https://huggingface.co/intfloat/multilingual-e5-small}
       {\texttt{intfloat/multilingual-e5-small}} \\
  \bottomrule
  \end{tabular}
  \caption{Encoders used by EviRerank (HF checkpoints).}
  \label{tab:cross-bi-encoders}
\end{table}

\paragraph{AEB hyperparameters.}
AEB applies ratio-based early stopping with threshold $\rho$ (Eq.~\ref{eq:keyb2_dynamic_stop}). For score normalization, we use \textsc{None} for BM25 and Min-Max normalization for neural selectors, computed over blocks \emph{within each document}. This design reflects that BM25 scores are derived from a fixed lexical matching function and are typically more comparable across blocks for the same query, whereas neural relevance scores can vary in scale and dynamic range across queries/documents, making the ratio criterion sensitive without normalization. We select $\rho$ on the dev set by sweeping $\{0.05, 0.10, 0.15,\ldots,0.65\}$ and then fix it for test. We also tune the minimum number of kept blocks $m$ (Eq.~\ref{eq:keyb2_dynamic_stop}) on the dev set by sweeping $\{2,3,4,5,6\}$, while other AEB parameters follow the default setting (e.g., maximum kept blocks).

\paragraph{BM25 settings.}
For English, we use Anserini defaults ($k_1{=}0.9$, $b{=}0.4$). For MLDR-zh, we apply Chinese word segmentation with jieba\footnote{\url{https://github.com/fxsjy/jieba}}; other settings follow English. For EviRerank$_{\text{BM25}}$ and KeyB models, IDF statistics are computed with \texttt{scikit-learn} on each corpus.

\section{Experimental Results}
\label{sec:experimentResults}

\begin{table}[h]
  \centering
  \caption{Results on TREC DL'19 document reranking, compared with sparse-attention rerankers and prior block-selection methods. Best results are in {\bf bold}. $^{\dagger}$ and $^{\ddagger}$ indicate statistically significant improvements under a paired two-sided t-test at $p<0.05$ over RankLLaMA and KeyB(BERT)$_{\text{BinB}}$, respectively. }
  \label{tab:dl19}
  \resizebox{0.77\linewidth}{!}{

\begin{tabular}{lll}
\toprule
\multicolumn{3}{r}{\textbf{TREC 2019 DL Track Document Reranking}} \\
\hline
Model   & NDCG@10 & MAP \\ 
\midrule
\multicolumn{3}{l}{\textit{\textbf{Baseline models}}} \\
BM25 & 0.488 & 0.234 \\
TKL      & 0.644 & 0.277 \\ 
PARADE        & 0.655 & 0.280 \\
\midrule
\multicolumn{3}{l}{\textit{\textbf{Sparse attention models}}} \\ 
Sparse-Transformer       & 0.634 & 0.257 \\ 
Longformer-QA      & 0.627 & 0.255 \\ 
Transformer-XH      & 0.646 & 0.256 \\
QDS-Transformer      & 0.667 & 0.278 \\
\midrule
\multicolumn{3}{l}{\textit{\textbf{Select blocks models}}} \\ 
IDCM  & 0.679 & 0.273 \\
KeyB(PARADE5)$_{BM25}$  & 0.672 & 0.280 \\ 
KeyB(PARADE5)$_{BinB}$  & 0.676 & 0.277 \\
KeyB(PARADE5)$_{BinB2}$  & 0.678 & 0.279 \\
KeyB(BERT)$_{BM25}$  & 0.683 & 0.281 \\
KeyB(BERT)$_{BinB}$  & 0.697 & 0.283 \\
\midrule
\multicolumn{3}{l}{\textit{\textbf{LLM}}} \\
RankLLaMA & 0.701 & 0.288 \\
RankLLaMA-prefix ($p{=}600$) & 0.692 & 0.281 \\
RankLLaMA-MaxP &0.643& 0.269 \\
RankLLaMA-AvgP &0.654&0.264 \\
\midrule
\multicolumn{3}{l}{\textit{\textbf{Ours: EviRerank (AEB+SA)}}} \\
EviRerank$_{\text{BM25}}$  & \textbf{0.744}$^{\dagger}$$^{\ddagger}$ & 0.302$^{\ddagger}$ \\
EviRerank$_{\text{cross}}$ & 0.743$^{\dagger}$$^{\ddagger}$ & \textbf{0.307}$^{\ddagger}$ \\
EviRerank$_{\text{bi}}$    & 0.743$^{\dagger}$$^{\ddagger}$ & 0.300$^{\ddagger}$ \\
\bottomrule
\end{tabular}

  }
\end{table}

\begin{table}[h]
  \centering
  \caption{Results on TREC DL'23 document reranking. Best results are in {\bf bold}.$^{\dagger}$ and $^{\ddagger}$ indicate statistically significant improvements under a paired two-sided t-test at $p<0.05$ over RankLLaMA and KeyB(BERT)$_{\text{BinB}}$, respectively.}
  \label{tab:dl23}
  \resizebox{0.75\linewidth}{!}{
  
\begin{tabular}{lll}
\toprule
\multicolumn{3}{r}{\textbf{TREC 2023 DL Track Document Reranking}} \\
\hline
Model   & NDCG@10 & MAP \\ 
\midrule
\multicolumn{3}{l}{\textit{\textbf{Baseline models}}} \\
BM25 & 0.295  & 0.105 \\
KeyB(BERT)$_{BM25}$  & 0.352 & 0.123 \\
KeyB(BERT)$_{BinB}$  & 0.364 & 0.128 \\
\midrule
\multicolumn{3}{l}{\textit{\textbf{LLM}}} \\
RankLLaMA & 0.386 & 0.133 \\
RankLLaMA-MaxP & 0.341& 0.112\\
RankLLaMA-AvgP & 0.349 & 0.117 \\
\midrule
\multicolumn{3}{l}{\textit{\textbf{Ours: EviRerank (AEB+SA)}}} \\
EviRerank$_{\text{BM25}}$  & 0.457$^{\dagger}$$^{\ddagger}$ & 0.154$^{\dagger}$$^{\ddagger}$ \\
EviRerank$_{\text{cross}}$ & \textbf{0.475}$^{\dagger}$$^{\ddagger}$ & \textbf{0.157}$^{\dagger}$$^{\ddagger}$ \\
EviRerank$_{\text{bi}}$    & 0.465$^{\dagger}$$^{\ddagger}$ & 0.156$^{\dagger}$$^{\ddagger}$ \\
\bottomrule
\end{tabular}

  }
\end{table}

\begin{table}[htbp]
\centering
\caption{Results on MLDR-zh. Best results are in {\bf bold}. $^{\dagger}$ and $^{\ddagger}$ indicate statistically significant improvements under a paired two-sided t-test at $p<0.05$ over RankLLaMA and KeyB(BERT)$_{\text{BinB}}$, respectively.}
\label{tab:mldr-zh}
\resizebox{0.8\linewidth}{!}{

\begin{tabular}{llll}
\toprule
Model & P@1 & MAP & NDCG@8 \\ 
\midrule
\multicolumn{4}{l}{\textit{\textbf{Baseline models}}} \\
BM25 & 0.201 & 0.259 & 0.277 \\ 
\midrule
\multicolumn{4}{l}{\textit{\textbf{BERT based models}}} \\
KeyB(BERT)$_{BM25}$ & 0.423 & 0.586 & 0.685 \\
KeyB(BERT)$_{BinB}$ & 0.804$^{\dagger}$ & 0.876$^{\dagger}$ & 0.907$^{\dagger}$ \\
\midrule
\multicolumn{4}{l}{\textit{\textbf{LLM}}} \\
RankLLaMA & 0.649 & 0.755 & 0.814 \\
RankLLaMA-MaxP & 0.374 & 0.554 & 0.661 \\
RankLLaMA-AvgP & 0.228 & 0.433 & 0.567 \\
\midrule
\multicolumn{4}{l}{\textit{\textbf{Ours: EviRerank (AEB+SA)}}} \\
EviRerank$_{\text{BM25}}$  & 0.909$^{\dagger}$$^{\ddagger}$ & 0.939$^{\dagger}$$^{\ddagger}$ & 0.954$^{\dagger}$$^{\ddagger}$ \\
EviRerank$_{\text{cross}}$ & 0.945$^{\dagger}$$^{\ddagger}$ & 0.964$^{\dagger}$$^{\ddagger}$ & 0.973$^{\dagger}$$^{\ddagger}$ \\
EviRerank$_{\text{bi}}$    & \textbf{0.951}$^{\dagger}$$^{\ddagger}$ & \textbf{0.968}$^{\dagger}$$^{\ddagger}$ & \textbf{0.976}$^{\dagger}$$^{\ddagger}$ \\
\bottomrule
\end{tabular}

}
\end{table}

\subsection{Dataset-wise Effectiveness Results}
\label{sec:results-dataset}

Tables~\ref{tab:dl19}--\ref{tab:mldr-zh} report the main pointwise results on DL'19, DL'23, and MLDR-zh. We additionally evaluate DL'22 as a held-out validation benchmark; its complete results are reported in Table~\ref{tab:dl22} of Appendix~\ref{app:dl22}. Significance is assessed by a paired two-sided $t$-test ($p \le 0.05$) on per-query metric scores. Superscripts $^{\dagger}$ and $^{\ddagger}$ denote statistically significant improvements over RankLLaMA and KeyB(BERT)$_{\text{BinB}}$, respectively.

\paragraph{TREC DL'19.}
The best configuration is \mbox{EviRerank (AEB+SA)}$_{\text{cross}}$ with \textbf{nDCG@10}~0.743 and \textbf{MAP}~0.307 (Table~\ref{tab:dl19}). This corresponds to clear improvements over the full-document LLM reranker RankLLaMA (0.701/0.288) and the strongest BERT-based block-selection baseline KeyB(BERT)$_{\text{BinB}}$ (0.697/0.283). Across selectors (BM25/bi/cross), EviRerank consistently outperforms RankLLaMA-MaxP/AvgP, showing that simple score pooling is inferior to explicitly selecting and composing evidence before reranking. Importantly, the gain is not explained by input shortening alone: the matched-budget RankLLaMA-prefix control reaches only 0.692 nDCG@10 and 0.281 MAP under the same $p_{\max}=600$ document-side cap. Moreover, all EviRerank (AEB+SA) variants are statistically significant ($p\le0.05$) over RankLLaMA ($^{\dagger}$) and KeyB(BERT)$_{\text{BinB}}$ ($^{\ddagger}$) on nDCG@10, and they are also significant over KeyB(BERT)$_{\text{BinB}}$ on MAP (Table~\ref{tab:dl19}).

\paragraph{TREC DL'23.}
On DL'23, the strongest result is obtained by \mbox{EviRerank (AEB+SA)}$_{\text{cross}}$, reaching \textbf{nDCG@10}~0.475 and \textbf{MAP}~0.157 (Table~\ref{tab:dl23}). Relative to RankLLaMA (0.386/0.133) and KeyB(BERT)$_{\text{BinB}}$ (0.364/0.128), this yields substantial gains on both metrics. Again, evidence construction (selection + composition) dominates aggregation-only baselines (MaxP/AvgP), reinforcing that allocating the input budget to high-utility evidence is more effective than pooling over block scores. All EviRerank (AEB+SA) variants yield statistically significant gains ($p\le0.05$) over both RankLLaMA ($^{\dagger}$) and KeyB(BERT)$_{\text{BinB}}$ ($^{\ddagger}$) on both metrics (Table~\ref{tab:dl23}).

\paragraph{TREC DL'22.}
As an additional held-out MS MARCO v2 document-ranking benchmark, DL'22 exhibits the same overall trend. Without DL'22-specific tuning, EviRerank$_{\text{cross}}$ achieves 0.4488 nDCG@10 and 0.1151 MAP, compared with 0.3845/0.1019 for full-document RankLLaMA. Complete results are reported in Table~\ref{tab:dl22} of Appendix~\ref{app:dl22}.

\paragraph{MLDR-zh.}
On MLDR-zh, the best configuration is \mbox{EviRerank (AEB+SA)}$_{\text{bi}}$, achieving the strongest scores across the reported metrics (Table~\ref{tab:mldr-zh}). Compared with RankLLaMA and KeyB(BERT)$_{\text{BinB}}$, EviRerank yields large improvements, highlighting its robustness for very long Chinese documents. Notably, the bi-encoder selector is particularly effective in this setting, likely because it provides a strong semantic prior under domain- and language-matched encoders. All EviRerank (AEB+SA) variants are statistically significant ($p\le0.05$) over RankLLaMA ($^{\dagger}$) and KeyB(BERT)$_{\text{BinB}}$ ($^{\ddagger}$) on all reported metrics (Table~\ref{tab:mldr-zh}). Interestingly, KeyB(BERT)$_{\text{BinB}}$ also significantly outperforms RankLLaMA on MLDR-zh ($^{\dagger}$), suggesting that strong BERT selectors can remain competitive for very long Chinese documents.

\paragraph{Cross-dataset takeaway (RQ2).}
EviRerank consistently improves over the controlled RankLLaMA-style baselines across DL'19, DL'22, DL'23, and MLDR-zh (Tables~\ref{tab:dl19}--\ref{tab:mldr-zh} and Table~\ref{tab:dl22}). Selector choice remains collection-dependent: BM25/cross-encoder variants are strongest on MS MARCO-style web documents, while the bi-encoder is strongest on MLDR-zh. Overall, evidence construction remains effective across datasets with different document-length and language profiles.

\subsection{Generality to Listwise LLM Reranking}
\label{sec:rankzephyr}

\begin{figure}[ht]
    \centering
    \includegraphics[width=\linewidth]{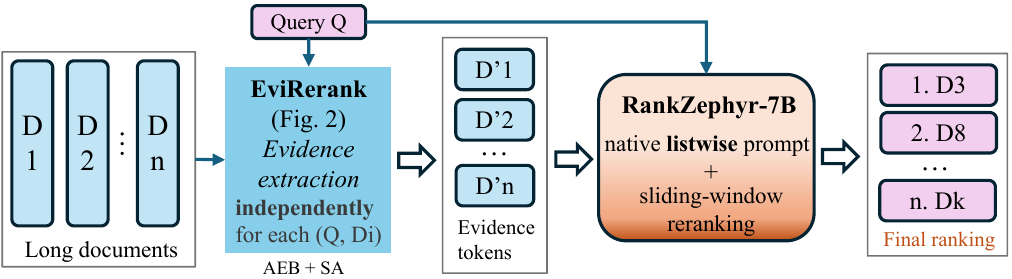}
    \caption{Applying EviRerank to RankZephyr-style listwise reranking. EviRerank independently constructs a compact representation for each candidate document, after which RankZephyr applies its native sliding-window listwise reranking protocol.}
    \label{fig:rankzephyr}
\end{figure}

To test whether EviRerank's gains are tied to the RankLLaMA-style pointwise scoring architecture, we apply the same evidence construction independently to every candidate document before RankZephyr-7B listwise reranking. Unlike Eq.~\ref{eq:keyb2_rsv}, no document-level linear scoring head is used; the resulting compact document representations are instead passed to RankZephyr's native listwise prompt and sliding-window reranking procedure. Figure~\ref{fig:rankzephyr} illustrates this interface, and Table~\ref{tab:rankzephyr} reports the matched-budget listwise reranking results.

\begin{table}[ht]
\centering
\scriptsize
\setlength{\tabcolsep}{3.5pt}
\renewcommand{\arraystretch}{1.02}
\caption{RankZephyr-7B listwise validation on TREC DL'19 and DL'23 under matched candidate sets, context budgets, and listwise protocols. Within each budget $p$, bold denotes the better result. Superscript $^{*}$ marks a significant improvement over matched full-document truncation ($p<0.05$, paired two-sided $t$-test).}
\label{tab:rankzephyr}
\resizebox{\columnwidth}{!}{
\begin{tabular}{lclcc}
\toprule
Dataset & $p$ & Method & nDCG@10 & MAP \\
\midrule
DL'19 & 150 & Full-doc trunc. & 0.6494 & 0.2691 \\
      &     & EviRerank       & \textbf{0.6859} & \textbf{0.2951}$^{*}$ \\
DL'19 & 300 & Full-doc trunc. & 0.6768 & 0.2769 \\
      &     & EviRerank       & \textbf{0.7263}$^{*}$ & \textbf{0.2910} \\
DL'19 & 600 & Full-doc trunc. & 0.6306 & \textbf{0.2660} \\
      &     & EviRerank       & \textbf{0.6349} & 0.2656 \\
\midrule
DL'23 & 150 & Full-doc trunc. & 0.4119 & 0.1451 \\
      &     & EviRerank       & \textbf{0.4441} & \textbf{0.1504} \\
DL'23 & 300 & Full-doc trunc. & 0.4271 & 0.1376 \\
      &     & EviRerank       & \textbf{0.4732}$^{*}$ & \textbf{0.1516}$^{*}$ \\
DL'23 & 600 & Full-doc trunc. & 0.3994 & \textbf{0.1322} \\
      &     & EviRerank       & \textbf{0.4012} & 0.1318 \\
\bottomrule
\end{tabular}
}
\end{table}

\noindent\textbf{Results.} EviRerank provides the clearest gains under tighter per-document budgets. At $p=300$, it improves RankZephyr from 0.6768 to 0.7263 nDCG@10 on DL'19 and from 0.4271 to 0.4732 on DL'23, with corresponding MAP gains. The same overall pattern also holds at $p=150$.

At $p=600$, however, the improvement largely saturates: nDCG@10 changes only marginally and MAP is essentially tied. This suggests that evidence construction is most useful when the per-document context budget is constrained, whereas its marginal benefit decreases once substantially more evidence can already be accommodated.

\noindent\textbf{Diagnostic controls.} Additional matched controls in Table~\ref{tab:rankzephyr-appendix}(b) of Appendix~\ref{app:rankzephyr-protocol} show that the improvement is not explained by arbitrary non-prefix block replacement: query-focused evidence is necessary, and AEB+SA performs best.

\subsection{Selector Choice and Effectiveness-Efficiency Trade-off (RQ3)}
\label{sec:rq3-selector}

\noindent\textbf{Which selector to use.} The optimal selector depends on the collection:
\begin{itemize}
\item \textbf{Cross-encoder} yields the highest effectiveness on MS~MARCO-style web documents (DL'19/DL'23), benefiting from fine-grained token interactions, at the cost of higher online scoring.
\item \textbf{Bi-encoder} provides a strong accuracy-efficiency balance when the encoder matches the language/domain (MLDR-zh). With cached block embeddings, query-time scoring is efficient.
\item \textbf{BM25} remains a strong efficiency-oriented option and is robust under lexical matching, especially when dense encoders are mismatched or unavailable.
\end{itemize}

\subsection{Does Summary Augmentation help? (RQ4)}
\label{sec:rq4-summaries}

Summary Augmentation (SA) adds a compact, query-agnostic global cue under the same total document-side cap. The controlled factorial experiments in Section~\ref{sec:ablation} show that SA improves effectiveness on both DL'19 and DL'23, while its combination with AEB yields the strongest results. A focused comparison against random-summary and no-summary controls is reported in Appendix~\ref{app:sa-diagnostic}.

\subsection{Efficiency (RQ5)}
\label{sec:rq5-efficiency}

\subsubsection{Inference Speed}

We measure end-to-end reranking latency under identical inference settings and report the wall-clock time required to rerank 100 candidate documents per query. The reported latency includes online block scoring, evidence construction, tokenization, and final LLM reranking. Fig.~\ref{fig:effecspeed} plots nDCG@10 against latency.

\noindent\textbf{Observations.} Full-document RankLLaMA is substantially slower (50.65s) while reaching 0.701 nDCG@10. EviRerank (AEB+SA) reaches 0.743--0.744 nDCG@10 at only 9.02--11.81s. Importantly, the matched-budget RankLLaMA-prefix control requires 17.90s but reaches only 0.692 nDCG@10, showing that the gain cannot be explained by shorter input alone. MaxP/AvgP are similarly efficient but substantially less effective (0.643--0.654 nDCG@10).

\subsubsection{Training Memory Requirements}

Appendix~\ref{sec:memuse} shows the training memory requirement comparisons, showing EviRerank offers a practical accuracy-resource trade-off for long-document reranking.

\section{Ablation Study}
\label{sec:ablation}

\subsection{Factorial Ablation: AEB and SA}
\label{sec:abl-factorial}

We conduct the same 2$\times$2 factorial ablation on DL'19 and DL'23 using the cross-encoder selector, isolating (i) AEB and (ii) SA. All runs use the same reranker, candidate set, hard $p_{\max}{=}600$ document-side cap, score normalization, and random seed; only AEB and SA are toggled.

\begin{table}[t]
\centering
\small
\setlength{\tabcolsep}{4pt}
\begin{tabular}{llcc}
\toprule
Dataset & Setting & nDCG@10 & MAP \\
\midrule
DL'19 & No-AEB/No-SA & 0.728 & 0.305 \\
DL'19 & AEB only     & 0.735 & 0.304 \\
DL'19 & SA only      & 0.739 & 0.305 \\
DL'19 & AEB+SA       & \textbf{0.743} & \textbf{0.307} \\
\midrule
DL'23 & No-AEB/No-SA & 0.465 & 0.155 \\
DL'23 & AEB only     & 0.470 & 0.156 \\
DL'23 & SA only      & 0.471 & 0.156 \\
DL'23 & AEB+SA       & \textbf{0.475} & \textbf{0.157} \\
\bottomrule
\end{tabular}
\caption{2$\times$2 factorial ablation of AEB and SA on DL'19 and DL'23 using the cross-encoder selector.}
\label{tab:ablation_2x2}
\end{table}

Table~\ref{tab:ablation_2x2} shows a consistent pattern across the two benchmarks. On DL'19, AEB improves nDCG@10 while reducing low-utility context, SA provides a larger effectiveness gain, and their combination performs best. The same ordering holds on DL'23: both components improve over the No-AEB/No-SA setting and AEB+SA again achieves the strongest result. Additional token-utilization and summary-control diagnostics are reported in Appendix~\ref{app:budget-diagnostics} and Appendix~\ref{app:sa-diagnostic}.

\section{Conclusion}
\label{sec:conclude}

We presented \textbf{EviRerank}, an evidence-guided framework for long-document reranking that constructs a compact document representation under a hard token cap before LLM scoring. Its \textbf{Adaptive Evidence Budgeting (AEB)} dynamically limits low-utility tail evidence, while \textbf{Summary Augmentation (SA)} adds a compact global cue under the same cap.

Across TREC DL'19, DL'22, DL'23, and MLDR-zh, EviRerank consistently improves over controlled full-document RankLLaMA-style reranking and strong block-selection baselines. Additional RankZephyr-7B experiments show that the same evidence-construction layer transfers to listwise reranking, with the largest gains under constrained per-document budgets. Together, the results show that long-document LLM reranking depends not only on which evidence to select, but also on how much evidence to retain under a limited context budget.

\section*{Limitations}

First, SA is lightweight and query-agnostic; richer query-conditioned or multi-facet global cues may provide additional benefits for documents with diverse intents. Second, although we evaluate four document-ranking benchmarks and validate both RankLLaMA-style pointwise and RankZephyr-style listwise reranking, the experiments do not cover all modern reranker architectures or retrieval domains. Finally, this work focuses on document reranking quality under a hard token budget. Evaluating whether the constructed evidence further improves downstream RAG or QA generation is complementary and left for future work.

\section*{Acknowledgments}
This work was partly funded by Institut Universitaire de France, IUF, and MIAI Cluster, ANR-23-IACL-0006. This work was also supported by the National Natural Science Foundation of China (No. 62376178).

\bibliography{custom}

\appendix

\section{Additional Experimental Results and Diagnostics}
\label{app:additional-validation}

\subsection{Additional DL'22 Validation}
\label{app:dl22}

We additionally evaluate EviRerank on TREC DL'22.
To keep DL'22 as a held-out validation benchmark, we reuse
the hyperparameters selected on DL'23 and perform no
DL'22-specific tuning.
Table~\ref{tab:dl22} reports the complete results.

\begin{table}[!ht]
\centering
\small
\begin{tabular}{lcc}
\toprule
Method & nDCG@10 & MAP \\
\midrule
BM25 & 0.2990 & 0.0800 \\
RankLLaMA & 0.3845 & 0.1019 \\
RankLLaMA-MaxP & 0.2248 & 0.0707 \\
RankLLaMA-AvgP & 0.3430 & 0.0984 \\
\midrule
EviRerank$_{\rm BM25}$ & 0.4307 & 0.1119 \\
EviRerank$_{\rm cross}$ & \textbf{0.4488} & \textbf{0.1151} \\
EviRerank$_{\rm bi}$ & 0.4412 & 0.1147 \\
\bottomrule
\end{tabular}
\caption{Additional validation on TREC DL'22 document reranking.
No DL'22-specific hyperparameter tuning is performed.}
\label{tab:dl22}
\end{table}

\subsection{RankZephyr Protocol and Diagnostic Controls}
\label{app:rankzephyr-protocol}

All RankZephyr experiments rerank the same official top-100
candidate set under a 4096-token context using the official
RankLLM prompt format. Full-document truncation and EviRerank
use identical listwise settings within each document budget.
Table~\ref{tab:rankzephyr-appendix}(a) summarizes the
budget-specific listwise settings, while
Table~\ref{tab:rankzephyr-appendix}(b) reports the matched
diagnostic controls on DL'19 at $p=300$.

\begin{table}[!ht]
\centering
\small

\textbf{(a) Listwise protocol}\\[2pt]
\begin{tabular}{ccc}
\toprule
$p$ & Window/stride & Evidence/SA budget \\
\midrule
150 & 20/10 & 120/30 \\
300 & 12/6  & 240/60 \\
600 & 6/3   & 480/120 \\
\bottomrule
\end{tabular}

\vspace{5pt}

\textbf{(b) Diagnostic controls on DL'19 at $p=300$}\\[2pt]
\begin{tabular}{lcc}
\toprule
Construction & nDCG@10 & MAP \\
\midrule
Full-document truncation & 0.6768 & 0.2769 \\
Random blocks & 0.6392 & 0.2584 \\
Fixed evidence (cross, no AEB/SA) & 0.7133 & 0.2861 \\
EviRerank (cross + AEB + SA) &
\textbf{0.7263} & \textbf{0.2910} \\
\bottomrule
\end{tabular}

\caption{RankZephyr-7B protocol and diagnostic controls.
Panel (a) reports the listwise settings for each per-document
budget; panel (b) compares evidence-construction variants on
DL'19 at $p=300$ under the same candidate set and listwise protocol.}
\label{tab:rankzephyr-appendix}
\end{table}

\subsection{Summary Augmentation Diagnostic}
\label{app:sa-diagnostic}

We compare the default SA cue against a random-block summary
and a no-summary control under the same maximum document-side cap.
Table~\ref{tab:abl-summary} reports the controlled comparison.

\begin{table}[!ht]
\centering
\small
\begin{tabular}{lc}
\toprule
Variant & nDCG@10 \\
\midrule
\textbf{EviRerank (SA, default)} & \textbf{0.739} \\
Random blocks summary & 0.731 \\
EviRerank (no SA) & 0.728 \\
\bottomrule
\end{tabular}
\caption{Summary augmentation diagnostic on DL'19 under the
same maximum document-side cap.}
\label{tab:abl-summary}
\end{table}

The default SA cue performs best, indicating that its benefit
cannot be explained by simply allocating the same budget to
additional query-focused or randomly selected blocks.

\subsection{Efficiency and Budget Diagnostics}
\label{app:budget-diagnostics}

\paragraph{Token utilization.}
We report the average document-side token usage of the
2$\times$2 ablation settings on DL'19. The measured length is
the tokenized document-side content after evidence composition
and truncation, before padding.

\begin{figure}[!htbp]
\centering
\includegraphics[width=\linewidth]{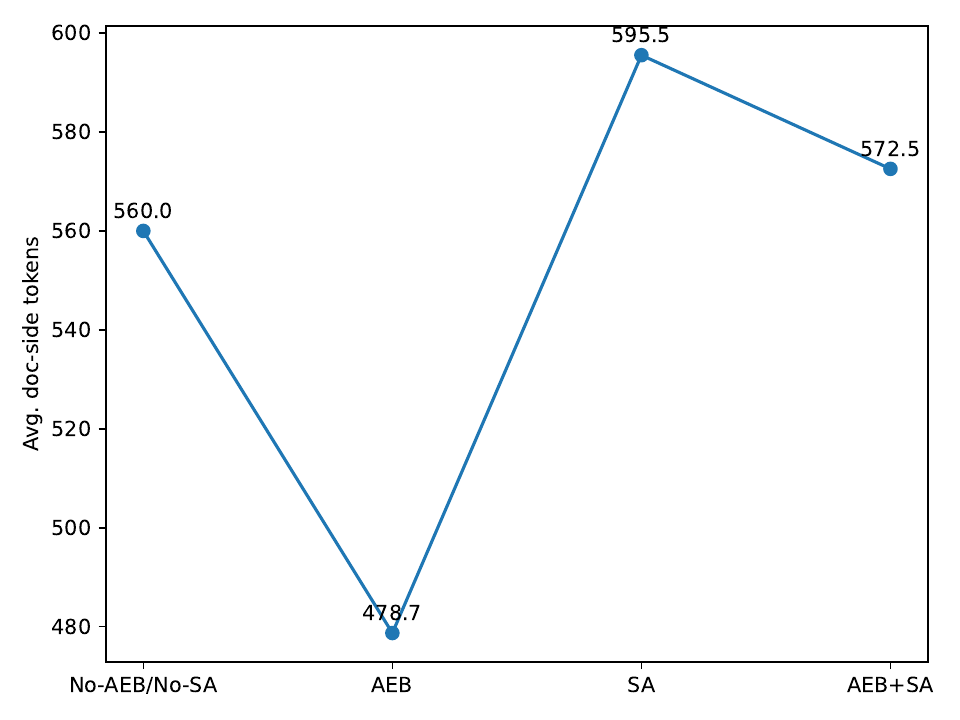}
\caption{Average document-side token usage of the
2$\times$2 ablations on DL'19. Values are annotated at each point.}
\label{fig:dl19_mean_tokens}
\end{figure}

As shown in Fig.~\ref{fig:dl19_mean_tokens}, AEB reduces the
average document-side length from 560.0 to 478.7 tokens
($-14.6\%$) without SA. When SA is enabled, the reduction is
smaller (595.5 to 572.5 tokens, $-3.9\%$), because SA already
consumes part of the shared token budget.

\paragraph{Budget sensitivity.}
We additionally vary the maximum document-side token budget
while keeping the other settings fixed.
Table~\ref{tab:budget-sensitivity} reports the results.

\begin{table}[!ht]
\centering
\small
\begin{tabular}{ccc}
\toprule
$p_{\max}$ & nDCG@10 & MAP \\
\midrule
400 & 0.731 & 0.306 \\
600 & 0.743 & 0.307 \\
800 & 0.744 & 0.307 \\
\bottomrule
\end{tabular}
\caption{Sensitivity to the document-side token cap on DL'19
using EviRerank$_{\rm cross}$. Other settings are fixed.}
\label{tab:budget-sensitivity}
\end{table}

As shown in Table~\ref{tab:budget-sensitivity}, performance
improves from 400 to 600 tokens and then largely saturates at
800 tokens, supporting $p_{\max}=600$ as a practical default.

\subsection{Training Memory Comparison}
\label{sec:memuse}

\begin{figure}[htbp]
  \centering
  \includegraphics[width=0.95\linewidth]{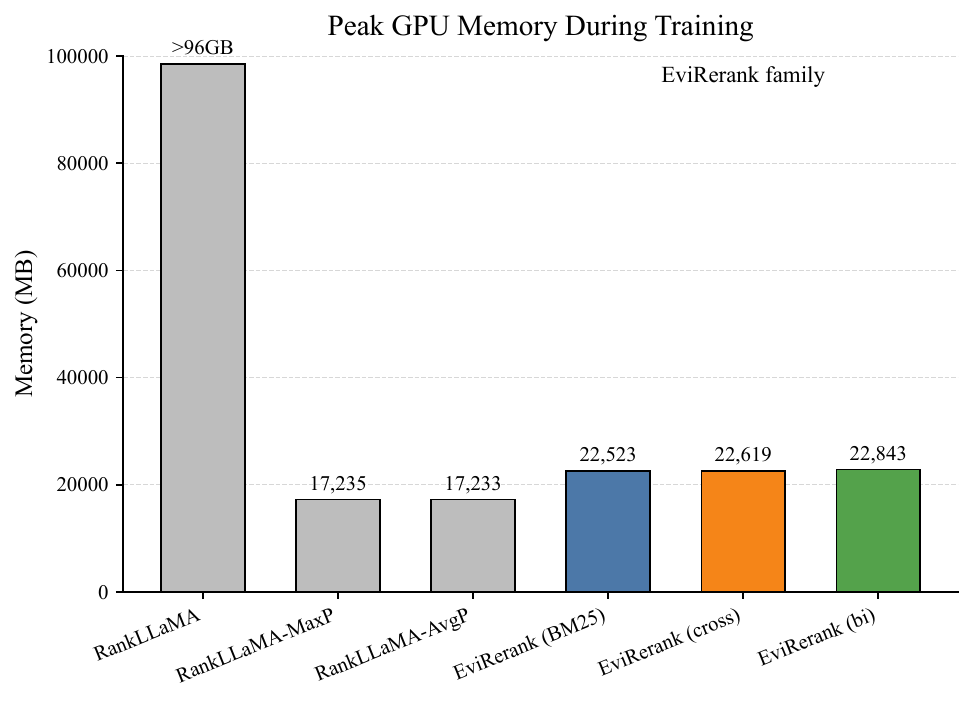}
  \caption{Peak GPU memory during training on DL19 (MB), measured on an NVIDIA H20 (96\,GB). 
  \textit{Note}: full-document RankLLaMA (4096-token input, batch size 1) exceeds 96\,GB and is reported as $>96$\,GB.}
  \label{fig:gpu}
  \vspace{-6pt}
\end{figure}

Fig.~\ref{fig:gpu} reports peak GPU memory usage during training on an NVIDIA H20 (96\,GB).
Full-document RankLLaMA, which consumes up to 4096 tokens per query-document pair, exceeds the 96\,GB capacity even with batch size 1; we therefore mark it as $>96$\,GB in the figure.
This highlights the high memory footprint of full-context training under long inputs.

In contrast, \textsc{EviRerank} trains within a much smaller memory budget by constructing a compact evidence context under a strict cap.
Across selectors, peak memory stays around 22-23\,GB.

Pooling-based baselines (RankLLaMA-MaxP/AvgP) are the most memory-efficient (17{,}235/17{,}233 MB),
but they are substantially less effective than \textsc{EviRerank} (Sec.~\ref{sec:results-dataset}).
Overall, budget-aware evidence construction offers a practical accuracy-resource trade-off for long-document reranking.

\section{Implementation and Reproducibility Details}
\label{app:implementation}

\subsection{Local Block Scorers}
\label{app:local-ranking}

This section provides the formal definitions and implementation details for the three local scorers used in EviRerank: BM25, cross-encoder, and bi-encoder.

\paragraph{BM25.}
Given a query $q$ and a block $blk$, the Retrieval Status Value (RSV) is:
\begin{equation*}
RSV_{\text{BM25}}(q,blk)=\sum_{w\in q\cap blk} IDF(w)\cdot s(w,blk),
\end{equation*}
\begin{equation*}
s(w,blk)=\frac{tf_w^{blk}}
{k_1\!\left(1-b+b\frac{l_{blk}}{l_{avg}}\right)+tf_w^{blk}} .
\end{equation*}
where $tf_w^{blk}$ denotes the frequency of term $w$ in block $blk$, $l_{blk}$ denotes the block length, and $l_{avg}$ denotes the average block length.

In this paper, we use the scikit-learn \cite{pedregosa2011scikit} package to calculate the IDF dictionary with IDF smoothing.
$IDF(w)$ is defined by:
\begin{equation}
\label{idf}
    IDF(w)=\log \frac{N+1}{df_{w}+1} +1, \nonumber
\end{equation}
where $N$ and $df_{w}$ are added one with default scikit-learn smooth IDF setting. The added 1 after the fraction, makes sure terms with zero IDF don't get suppressed entirely.
Besides, $l_{blk}$ is the length of block, $l_{avg}$ the average length of the blocks in $d$, and $k_1$ and $b$ two hyperparameters of BM25. 

For Chinese documents, word $w$ is recognized using the Jieba\footnote{\url{https://github.com/fxsjy/jieba}} Chinese text segmentation tool to obtain meaningful terms.

\paragraph{Cross-encoder (interaction).}
For each block $b_i$, a pretrained encoder (e.g., BERT) takes concatenated query and block tokens:
\[
\begin{aligned}
b_i^{cls} &= \mathrm{PLM}\big([CLS], q\_tokens,\\
           &\qquad [SEP], b_i\_tokens\big),
\end{aligned}
\]
where $b_i^{cls}$ is the [CLS] embedding. A linear layer maps it to a relevance score:
\[
RSV(q,b_i) = W b_i^{cls},
\]
with $W$ the learnable weights. Because blocks are short, this remains efficient while retaining rich interactions.

\paragraph{Bi-encoder (dense retrieval).}
A shared encoder independently maps queries and blocks:
\begin{equation*}
\begin{aligned}
\mathbf{q}^{\mathrm{cls}} &= \mathrm{PLM}\big([\mathrm{CLS}],\, \mathbf{q}_{\mathrm{tok}}\big),\\
\mathbf{b}_i^{\mathrm{cls}} &= \mathrm{PLM}\big([\mathrm{CLS}],\, \mathbf{b}_{i,\mathrm{tok}}\big).
\end{aligned}
\end{equation*}

Local relevance is computed as cosine similarity:
\begin{equation*}
RSV(q,b_i)=\cos\!\left(\mathbf{q}^{\mathrm{cls}},\,\mathbf{b}_i^{\mathrm{cls}}\right).
\end{equation*}
Block embeddings can be precomputed offline, making this approach extremely efficient at runtime.

\subsection{Dataset Statistics}
\label{app:dataSetStat}
Table~\ref{tab:datastesStats} summarizes genres, corpus sizes, test query counts, and average document lengths (tokenized by LLaMA2). 
\begin{table*}[htbp]
  \caption{Datasets statistics (with LLaMA2-7B tokenizer).}
  \centering
  \scalebox{0.8}{
  
\begin{tabular}{@{}lllll@{}}
\toprule
Dataset
&genre 
&\#documents &\#test query &Avg. Num. of Tokens \\ \midrule
TREC DL'19 doc (MS MARCO) & Web Documents &3,213,835& 43 & 1958\\
TREC DL'22 doc (MS MARCO v2) & Web Documents &11,959,635& 76 & 3782 \\
TREC DL'23 doc (MS MARCO v2) & Web Documents &11,959,635& 82 & 3782 \\
MLDR-zh &Wikipedia + Wudao&200,000& 800 &  12899\\
 \bottomrule
\end{tabular}

  }
  \label{tab:datastesStats}
\end{table*}

\subsection{Baseline Details}
\label{sec:moreBaselines}
We compare against competitive first-stage and reranking systems. Baseline sets align with prior work while remaining consistent across datasets where applicable.

\paragraph{DL'19 baselines (following \cite{li2023power}).}
\begin{itemize}
  \item \textit{Traditional IR}: BM25 (Anserini) \cite{yang2018anserini}.
  \item \textit{Neural IR}: TKL \cite{hofstatter2020local}, PARADE \cite{li2020parade}, Sparse-Transformer \cite{child2019}, Longformer-QA \cite{beltagy2020longformer}, Transformer-XH \cite{Zhao2020Transformer-XH:}, QDS-Transformer \cite{jiang2020long}.
  \item \textit{Key-block selection}: IDCM \cite{DBLP:conf/sigir/HofstatterMZCH21}, KeyB(PARADE5) and KeyB \cite{li2023power}.
  \item \textit{LLM reranker}: RankLLaMA \cite{ma2024fine} and our segment-level variants RankLLaMA-MaxP / RankLLaMA-AvgP (defined below).
\end{itemize}

\paragraph{DL'22 / DL'23 / MLDR-zh baselines.}
\begin{itemize}
  \item \textit{BM25}: Anserini for English; for MLDR-zh, we apply Chinese word segmentation.
  \item \textit{KeyB family}: KeyB with BM25-based / bi-encoder-based selection; English uses BERT and MLDR-zh uses Chinese BERT-style encoders.
  \item \textit{LLM reranker}: RankLLaMA and our MaxP / AvgP variants.
\end{itemize}

\subsection{Training Data and Evaluation Metrics}
\label{app:trainData}
\paragraph{Training data.}
We adopt standard triplet construction:
\begin{itemize}
  \item \textbf{DL’19}: 200K triplets from MS~MARCO v1 (positives from qrels; negatives from top-100).
  \item \textbf{DL’22/DL’23}: 200K triplets from
MS~MARCO v2 (same protocol).
  \item \textbf{MLDR-zh}: 10K labeled queries with one positive and one negative each.
\end{itemize}

\paragraph{Metrics.}
We follow community practice per dataset:
\begin{itemize}
  \item \textbf{DL’19/DL’22/DL’23}: NDCG@10, MAP.
  \item \textbf{MLDR-zh}: The test file for each query
  has one positive document together with seven negative
  documents. We therefore report P@1, MAP, and NDCG@8.
\end{itemize}

\section{Attention Analysis: Why Block Selection Still Matters}
\label{sec:attentionMap}

With the advent of PLMs like BERT, IR systems have seen substantial improvements in document ranking accuracy. Among these, re-ranking models, often referred to as cross-encoders, harness the power of fine-tuned PLMs or decoder-only LLMs for downstream IR tasks.
Despite the impressive performance of decoder-only LLMs like RankLLaMA, the inner workings of these LLMs, specifically how they assess and rank the relevance of passages, remain underexplored. 

\begin{figure*}[t]
\centering
\begin{subfigure}[t]{0.225\linewidth}
  \centering
  \includegraphics[width=\linewidth]{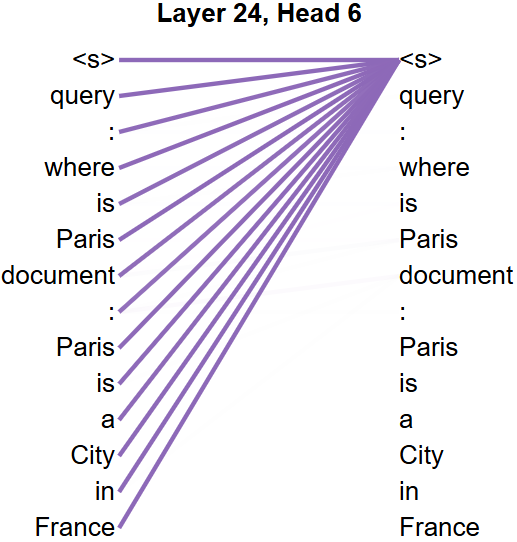}
  \caption{Delimiter heads}
  \label{fig1-a}
\end{subfigure}\hfill
\begin{subfigure}[t]{0.225\linewidth}
  \centering
  \includegraphics[width=\linewidth]{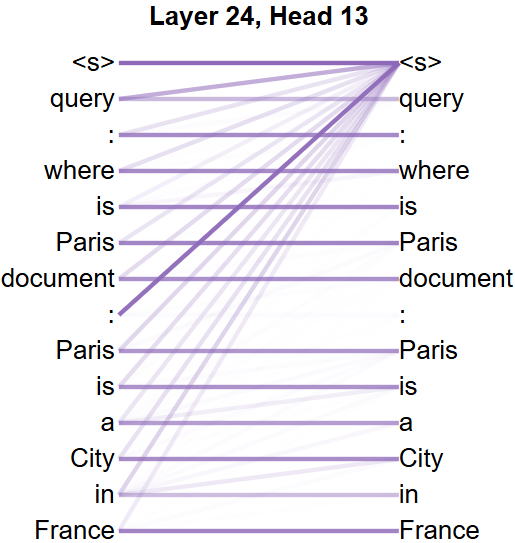}
  \caption{Local heads}
  \label{fig1-b}
\end{subfigure}\hfill
\begin{subfigure}[t]{0.225\linewidth}
  \centering
  \includegraphics[width=\linewidth]{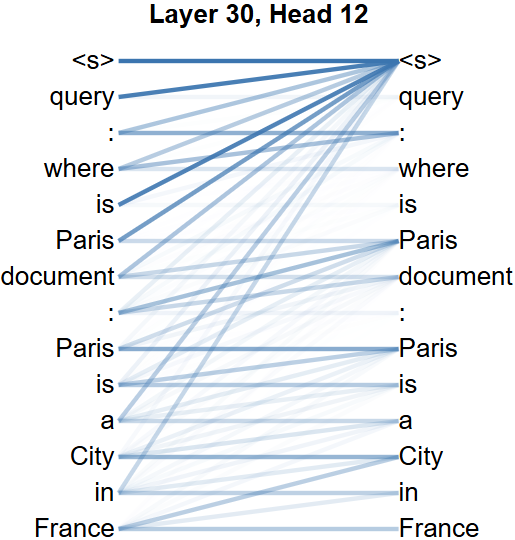}
  \caption{Broad heads}
  \label{fig1-c}
\end{subfigure}\hfill
\begin{subfigure}[t]{0.225\linewidth}
  \centering
  \includegraphics[width=\linewidth]{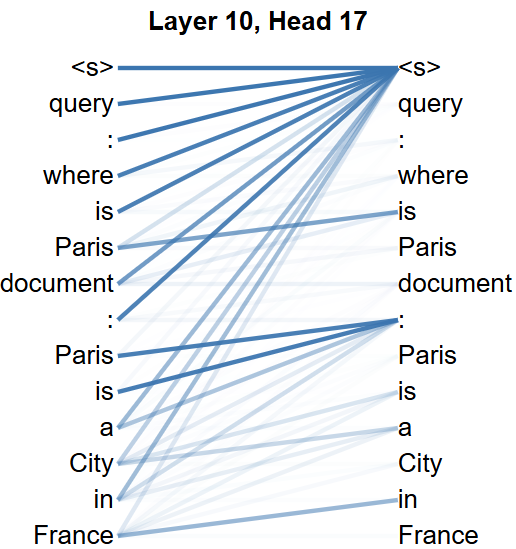}
  \caption{Broad heads}
  \label{fig1-d}
\end{subfigure}
\caption{Examples of attention heads that focus on delimiters, local context, or attend broadly.}
\label{fig:attendToS}
\vspace{-2mm}
\end{figure*}

\begin{figure*}[t]
\centering
\begin{subfigure}[t]{0.32\linewidth}
  \centering
  \includegraphics[width=\linewidth]{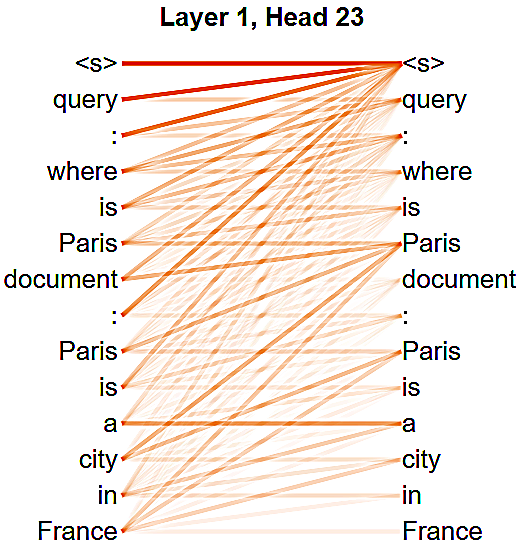}
  \caption{Overall map}
\end{subfigure}\hfill
\begin{subfigure}[t]{0.32\linewidth}
  \centering
  \includegraphics[width=\linewidth]{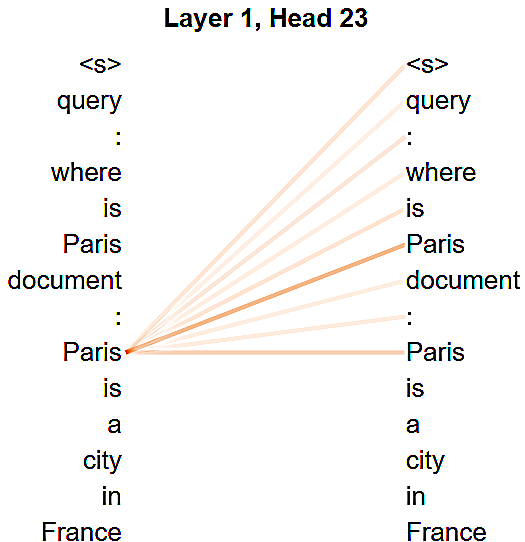}
  \caption{Token view \#1}
\end{subfigure}\hfill
\begin{subfigure}[t]{0.32\linewidth}
  \centering
  \includegraphics[width=\linewidth]{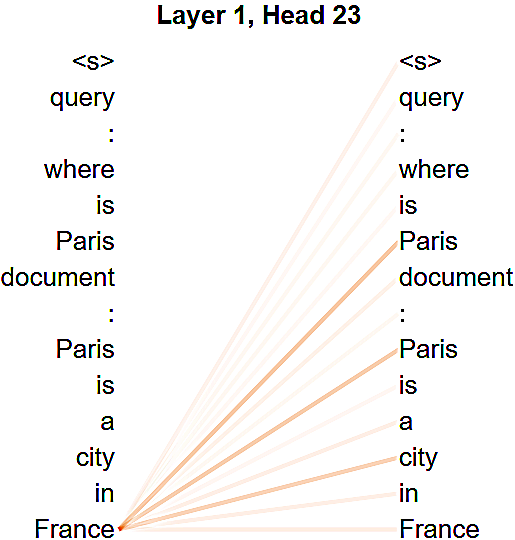}
  \caption{Token view \#2}
\end{subfigure}

\vspace{1mm}

\begin{subfigure}[t]{0.32\linewidth}
  \centering
  \includegraphics[width=\linewidth]{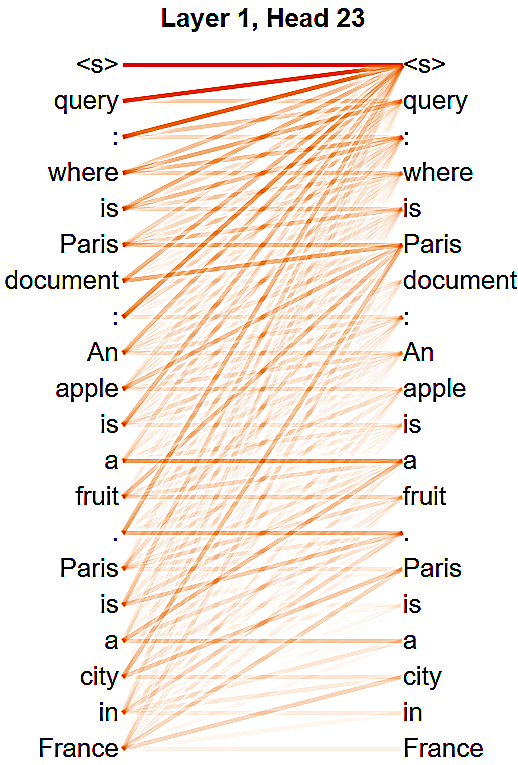}
  \caption{Overall (noise)}
\end{subfigure}\hfill
\begin{subfigure}[t]{0.32\linewidth}
  \centering
  \includegraphics[width=\linewidth]{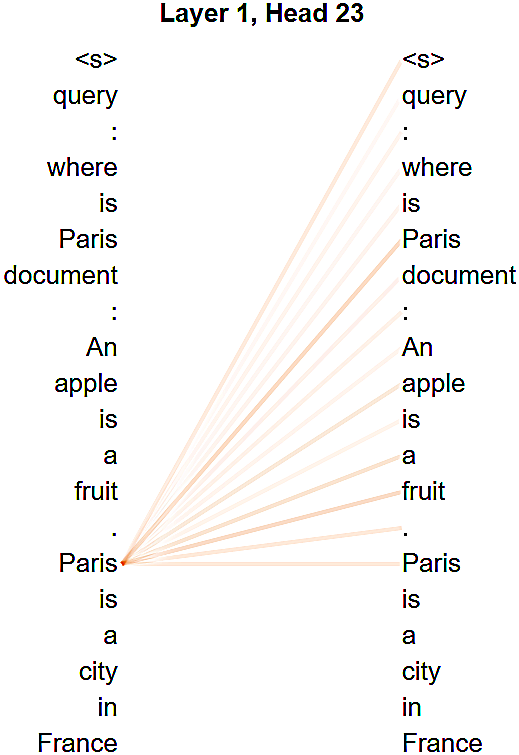}
  \caption{Token \#1 (noise)}
  \label{fig:attend1-23-e}
\end{subfigure}\hfill
\begin{subfigure}[t]{0.32\linewidth}
  \centering
  \includegraphics[width=\linewidth]{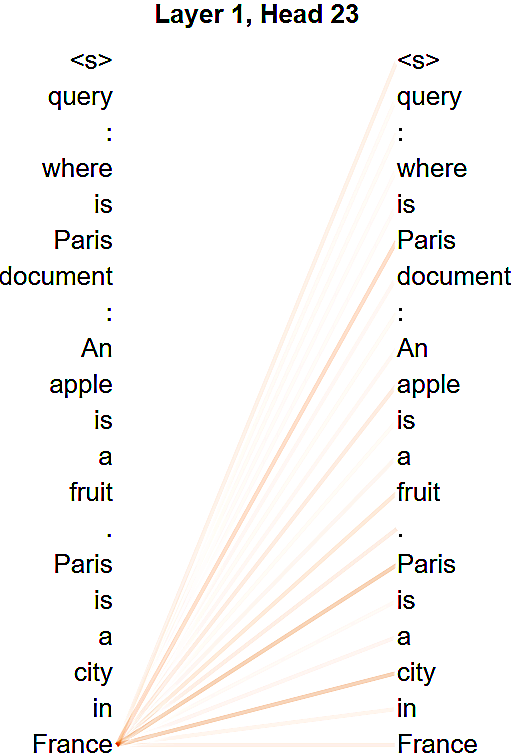}
  \caption{Token \#2 (noise)}
  \label{fig:attend1-23-f}
\end{subfigure}

\caption{Attention maps of Layer~1 Head~23 on a clean document (top) and with appended noise (bottom).}
\label{fig:attend1-23}
\vspace{-2mm}
\end{figure*}

\begin{figure*}[t]
\centering
\begin{subfigure}[t]{0.32\linewidth}
  \centering
  \includegraphics[width=\linewidth]{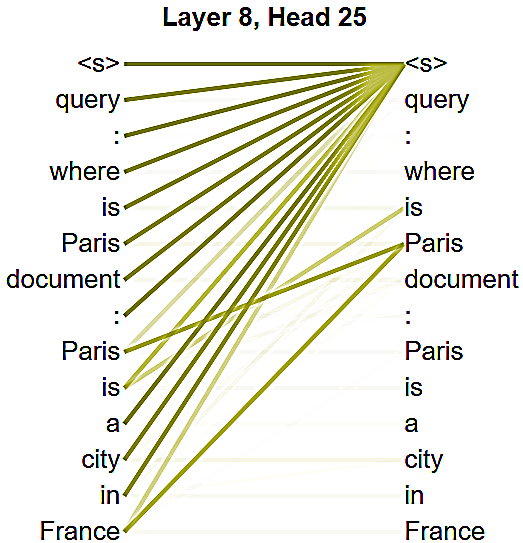}
  \caption{Overall map}
\end{subfigure}\hfill
\begin{subfigure}[t]{0.32\linewidth}
  \centering
  \includegraphics[width=\linewidth]{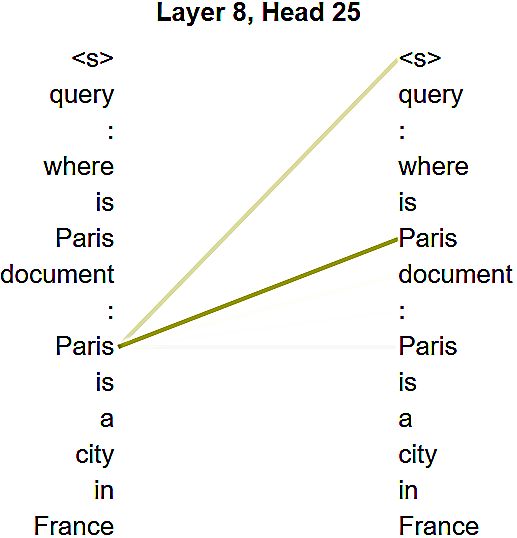}
  \caption{Token view \#1}
\end{subfigure}\hfill
\begin{subfigure}[t]{0.32\linewidth}
  \centering
  \includegraphics[width=\linewidth]{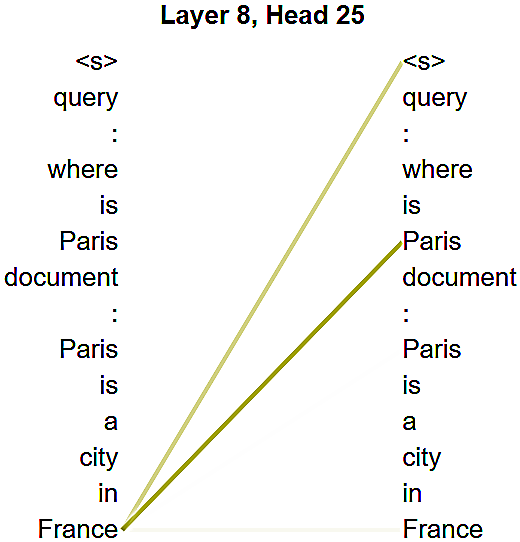}
  \caption{Token view \#2}
\end{subfigure}

\vspace{1mm}

\begin{subfigure}[t]{0.32\linewidth}
  \centering
  \includegraphics[width=\linewidth]{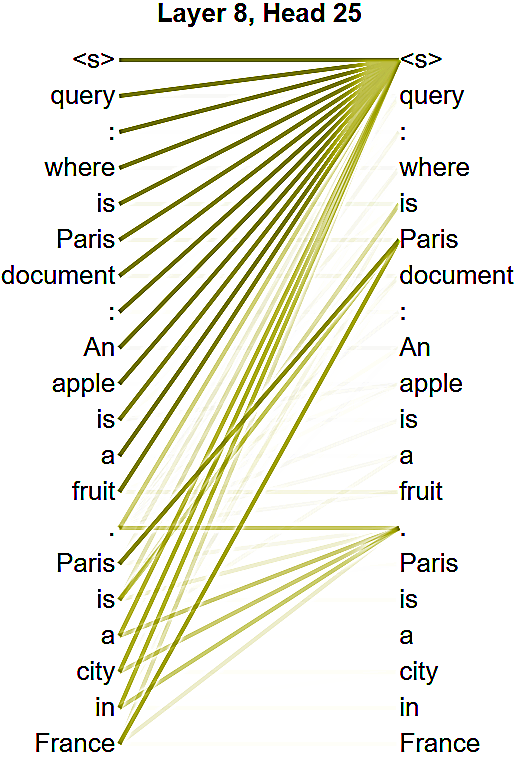}
  \caption{Overall (noise)}
\end{subfigure}\hfill
\begin{subfigure}[t]{0.32\linewidth}
  \centering
  \includegraphics[width=\linewidth]{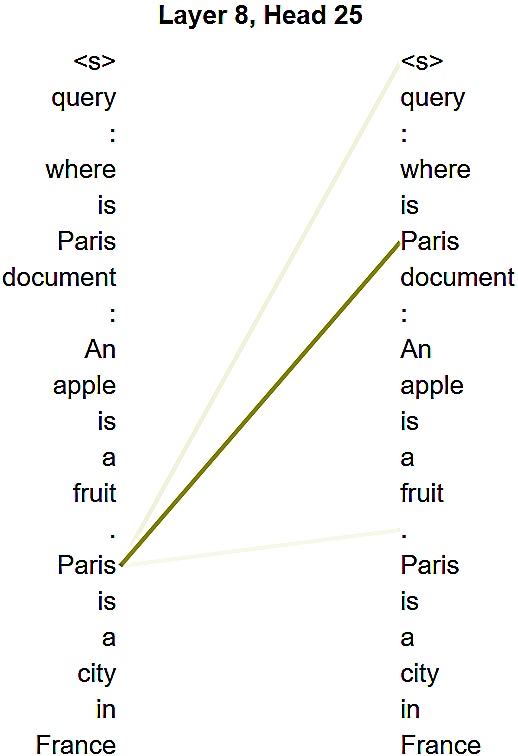}
  \caption{Token \#1 (noise)}
  \label{fig:attend-8-25-e}
\end{subfigure}\hfill
\begin{subfigure}[t]{0.32\linewidth}
  \centering
  \includegraphics[width=\linewidth]{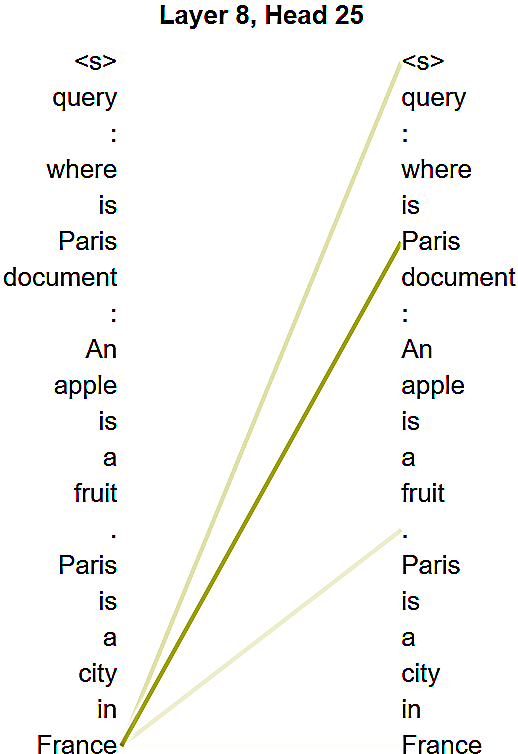}
  \caption{Token \#2 (noise)}
  \label{fig:attend-8-25-f}
\end{subfigure}

\caption{Attention maps of Layer~8 Head~25 on a clean document (top) and with appended noise (bottom).}
\label{fig:attend-8-25}
\vspace{-2mm}
\end{figure*}

\begin{figure*}[t]
\centering
\begin{subfigure}[t]{0.32\linewidth}
  \centering
  \includegraphics[width=\linewidth]{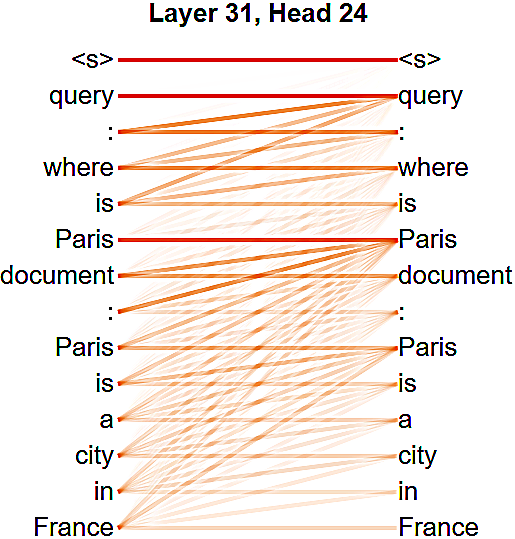}
  \caption{Overall map}
\end{subfigure}\hfill
\begin{subfigure}[t]{0.32\linewidth}
  \centering
  \includegraphics[width=\linewidth]{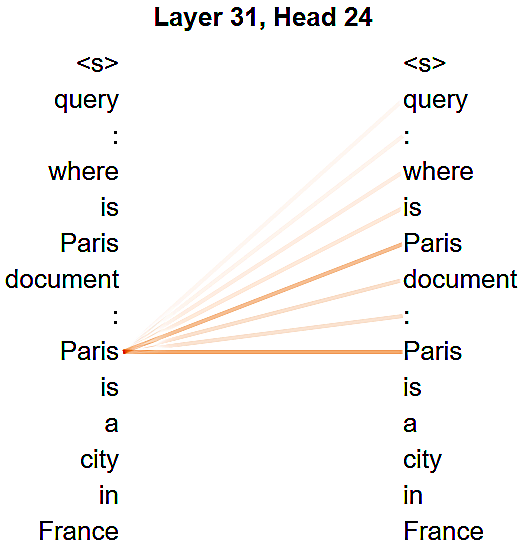}
  \caption{Token view \#1}
\end{subfigure}\hfill
\begin{subfigure}[t]{0.32\linewidth}
  \centering
  \includegraphics[width=\linewidth]{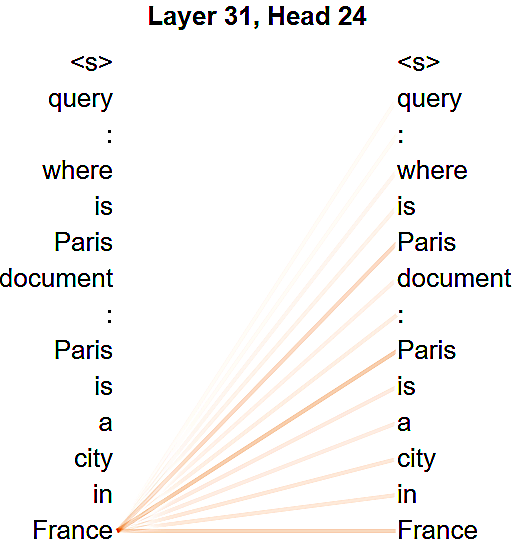}
  \caption{Token view \#2}
\end{subfigure}

\vspace{1mm}

\begin{subfigure}[t]{0.32\linewidth}
  \centering
  \includegraphics[width=\linewidth]{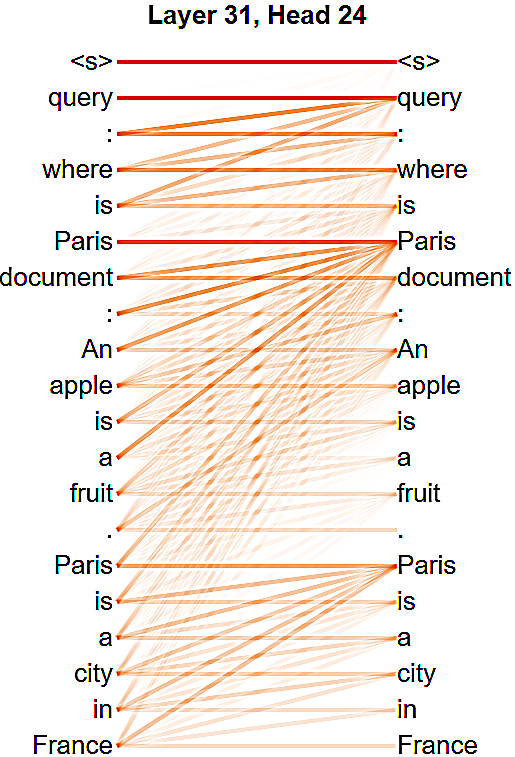}
  \caption{Overall (noise)}
\end{subfigure}\hfill
\begin{subfigure}[t]{0.32\linewidth}
  \centering
  \includegraphics[width=\linewidth]{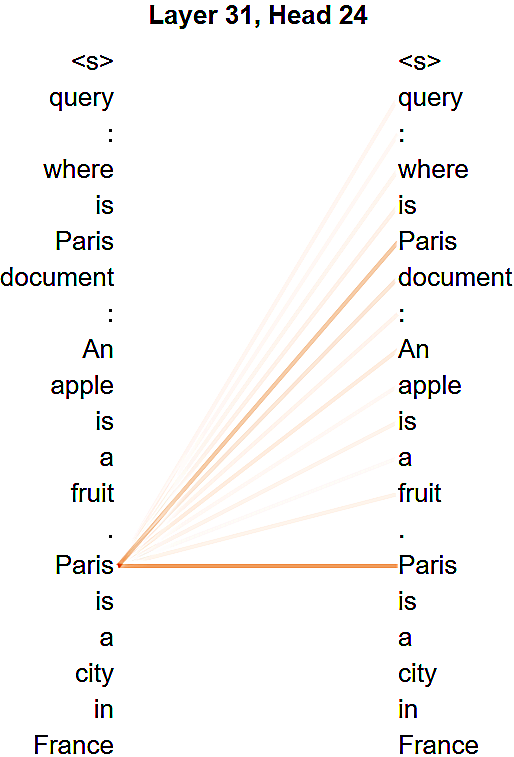}
  \caption{Token \#1 (noise)}
  \label{fig:attend-31-24-e}
\end{subfigure}\hfill
\begin{subfigure}[t]{0.32\linewidth}
  \centering
  \includegraphics[width=\linewidth]{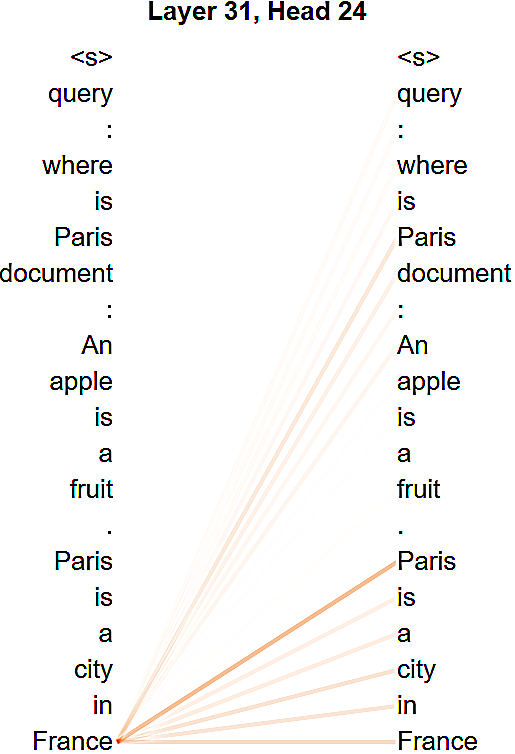}
  \caption{Token \#2 (noise)}
  \label{fig:attend-31-24-f}
\end{subfigure}

\caption{Attention maps of Layer~31 Head~24 on a clean document (top) and with appended noise (bottom).}
\label{fig:attend-31-24}
\vspace{-2mm}
\end{figure*}

\subsection{Attention Heatmaps of Specific Examples}
\label{sec:attention-simple-example}
To shed light on this and answer \textbf{RQ1}, we propose first analyzing the attention heatmaps of the RankLLaMA model with several specific examples, which has been fine-tuned on the MSMARCO passage ranking dataset\footnote{available at: \url{https://huggingface.co/castorini/rankLlama-v1-7b-lora-passage}}. 
\citet{clark2019does} propose a similar investigation for BERT and find a significant portion of BERT's attention is focused on the delimiter token, and certain attention heads align well with linguistic features such as syntax and coreference. 
However, recent LLMs are uni-directional and decoder-only, especially for the IR focused LLMs, and may display a different behavior regarding how tokens attend to each other.
We used the BertViz tool \cite{vig-2019-multiscale} to explore these attention patterns in the decoder-only LLM RankLLaMA when processing various query-passage pairs.

We begin with a simple and arbitrary text pair: the query text is "where is Paris", and the document text is "Paris is a city in France". With the format "query: [query] document: [document]",
after Llama2 tokenizer, the input tokens are: `<s>', `\_query', `:', `\_where', `\_is', `\_Paris', `\_document', `:', `\_Paris', `\_is', `\_a', `\_City', `\_in', `\_France'.
We check several attention heads which are shown in Fig.~\ref{fig:attendToS}, Fig.~\ref{fig:attend1-23}, Fig.~\ref{fig:attend-8-25} and Fig.~\ref{fig:attend-31-24}. 
To help showing weak attention weights, the last three figures are sharpened.
Our findings are summarized in the following points:
\begin{itemize}
    
    \item Attention to delimiter, current and broad tokens:
    Similar to \cite{clark2019does} in BERT, we observe that a substantial portion of attention is directed towards the delimiter token <s>, e.g., Fig.~\ref{fig1-a}.  \citet{clark2019does} speculate that attention over the <s> delimiter tokens might be used as a sort of “no-op”, and we conjecture that the attention over the beginning <s> is also "no-op" for IR scenario. Besides, we also observe a small amount of attention is directed towards the current token (Fig. \ref{fig1-b}), and that some attention heads attend broadly over all tokens (Fig. \ref{fig1-c}, Fig. \ref{fig1-d}), which is consistent with \cite{clark2019does}.

\item Attention focusing on relevant tokens: 
Beyond delimiter-focused or self-focused heads, we observe several attention heads that directly capture semantic relations between relevant tokens in the query and document. 
In particular, as shown in each above three subfigures in Fig.~\ref{fig:attend1-23}, Fig.~\ref{fig:attend-8-25}, and Fig.~\ref{fig:attend-31-24}, these relevance-focused heads consistently highlight key cross-token alignments:
for head 23 in layer 1, tokens such as ``Paris'' and ``France'' in the document primarily attend to corresponding relevant tokens in both query and document segments;
for head 25 in layer 8, tokens ``Paris'' and ``France'' focus strongly on the query token ``Paris'';
for head 24 in layer 31, similar alignment between query and document relevance tokens is also preserved.
These patterns suggest that a subset of attention heads are able to capture fine-grained token-level relevance signals, forming direct associations between query intents and document content—an essential mechanism underlying accurate relevance estimation.

\item Attention with irrelevant information:  
To investigate the effect of irrelevant tokens, we insert unrelated content such as ``An apple is a fruit'' into the document. This setting is illustrated in the three subfigures in Fig.~\ref{fig:attend1-23}, Fig.~\ref{fig:attend-8-25}, and Fig.~\ref{fig:attend-31-24}. The attention head 25 in layer 8 (Fig.~\ref{fig:attend-8-25-e}) effectively suppresses the irrelevant content, maintaining strong attention on the relevant query token ``Paris.'' This head appears to filter out the noise (e.g., apple) and prioritize query-relevant content.

In contrast, head 23 in layer 1 (Fig.~\ref{fig:attend1-23-e}) continues to exhibit broad attention across all tokens. Although the attention may be weak, it reveals that the head does not clearly distinguish between relevant and irrelevant content. Similarly, head 24 in layer 31 (Fig.~\ref{fig:attend-31-24-e}) shows that the token ``Paris'' in the document attends not only to the query token ``Paris'' but also weakly to irrelevant tokens such as ``apple,'' ``is,'' and ``a.''

Importantly, except for head 25 in layer 8, \textbf{the introduction of irrelevant tokens leads to a noticeable reduction in the attention weight toward the query token ``Paris.''} For instance, in Fig.~\ref{fig:attend1-23-f} and Fig.~\ref{fig:attend-31-24-f}, the attention from the final document token ``France'' to the query token ``Paris'' becomes weaker after inserting irrelevant information.

To support this observation quantitatively, Table~\ref{tab:attentionWeightExample} presents the measured attention scores from ``France'' to ``Paris'' in the query across the three heads. All heads exhibit decreased attention scores following the insertion of irrelevant tokens. Notably, while the decline for Layer 8 Head 25 is mild, the other two heads show substantial drops. This degradation may impair the model’s ability to correctly assess the relevance between document and query, which is particularly critical in information retrieval scenarios.
These irrelevant tokens will inevitably result in noisy information in the representation of the last token, which is the basic building block for computing the relevance score for RankLLaMA.

\end{itemize}

\begin{table*}[t]
\centering
\caption{The attention weights between ``France'' to ``Paris'' in the query when inserting noise tokens.}
\label{tab:attentionWeightExample}
\scalebox{0.95}{
\begin{tabular}{llll}
\toprule
&Layer 1 Head 23 (Fig~\ref{fig:attend1-23}) &Layer 8 Head 25 (Fig~\ref{fig:attend-8-25}) & Layer 31 Head 24 (Fig~\ref{fig:attend-31-24})\\
no noise tokens &0.2097 &0.6500 &0.1296 \\
with noise tokens &0.1154 (↓44.97\%) &0.6492 (↓0.12\%) &0.0858 (↓33.80\%) \\
\bottomrule
\end{tabular}

}
\end{table*}

\begin{figure}[htbp]
  \begin{subfigure}[t]{0.95\linewidth}
      \centering
      \includegraphics[width=\linewidth]{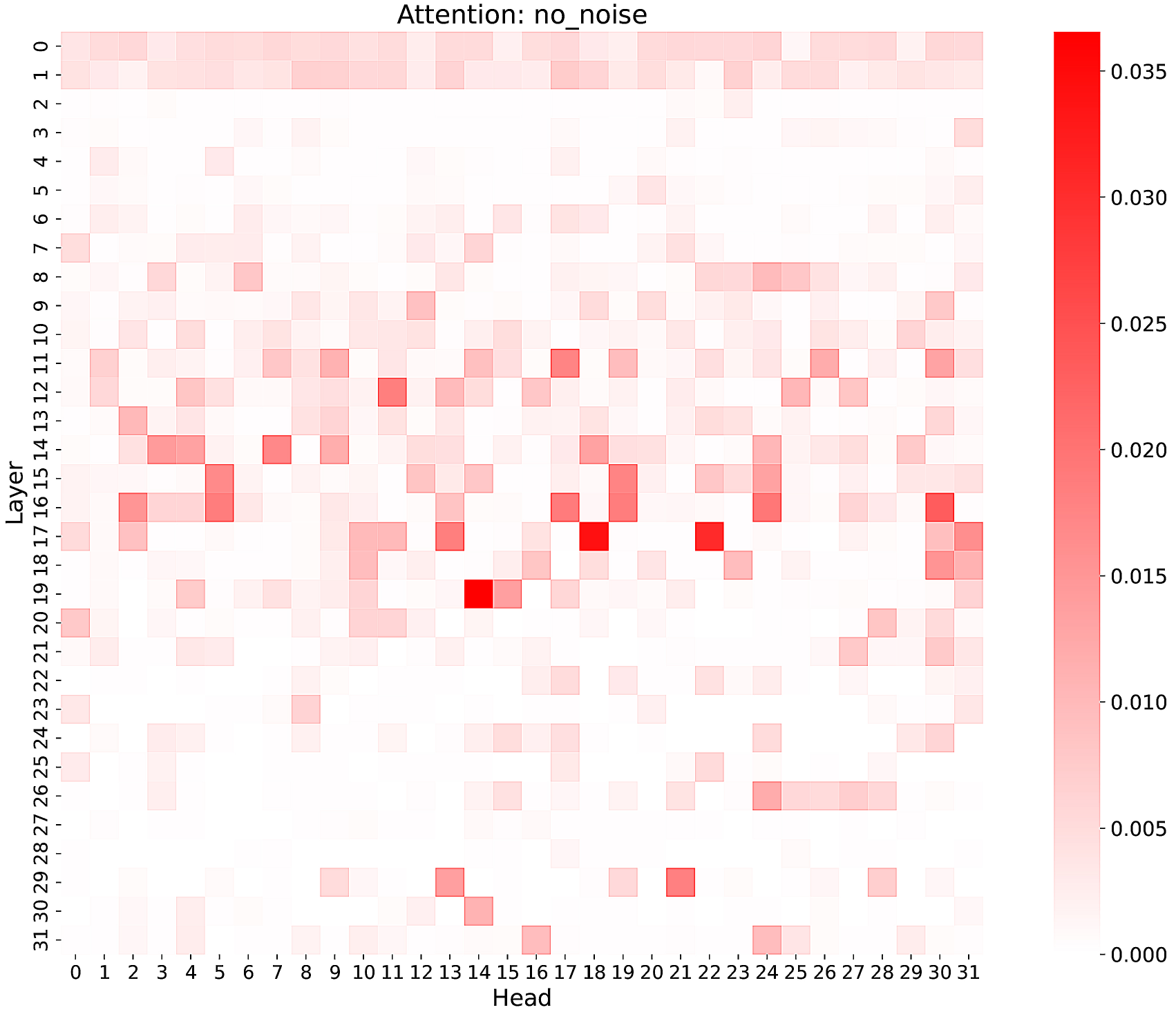}
      \caption{The overall attention heatmaps for relevant documents (without inserting noise)}
      \label{fig-overallAtten-a}
  \end{subfigure}%
  
  \begin{subfigure}[t]{0.95\linewidth}
      \centering
      \includegraphics[width=\linewidth]{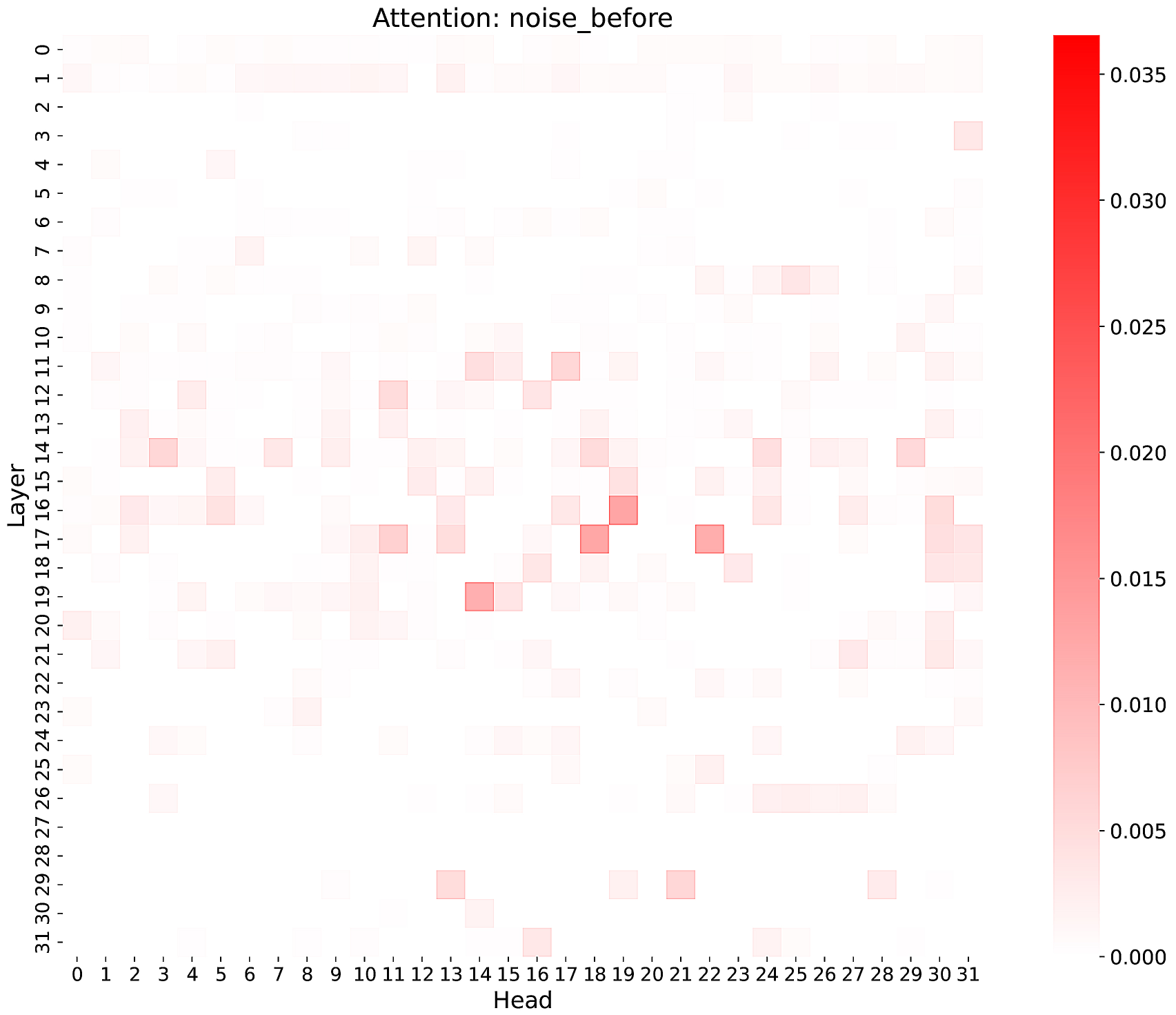}
      \caption{The overall attention heatmaps when inserting noise before relevant documents}
      \label{fig-overallAtten-b}
  \end{subfigure}%
  
  \begin{subfigure}[t]{0.95\linewidth}
      \centering
      \includegraphics[width=\linewidth]{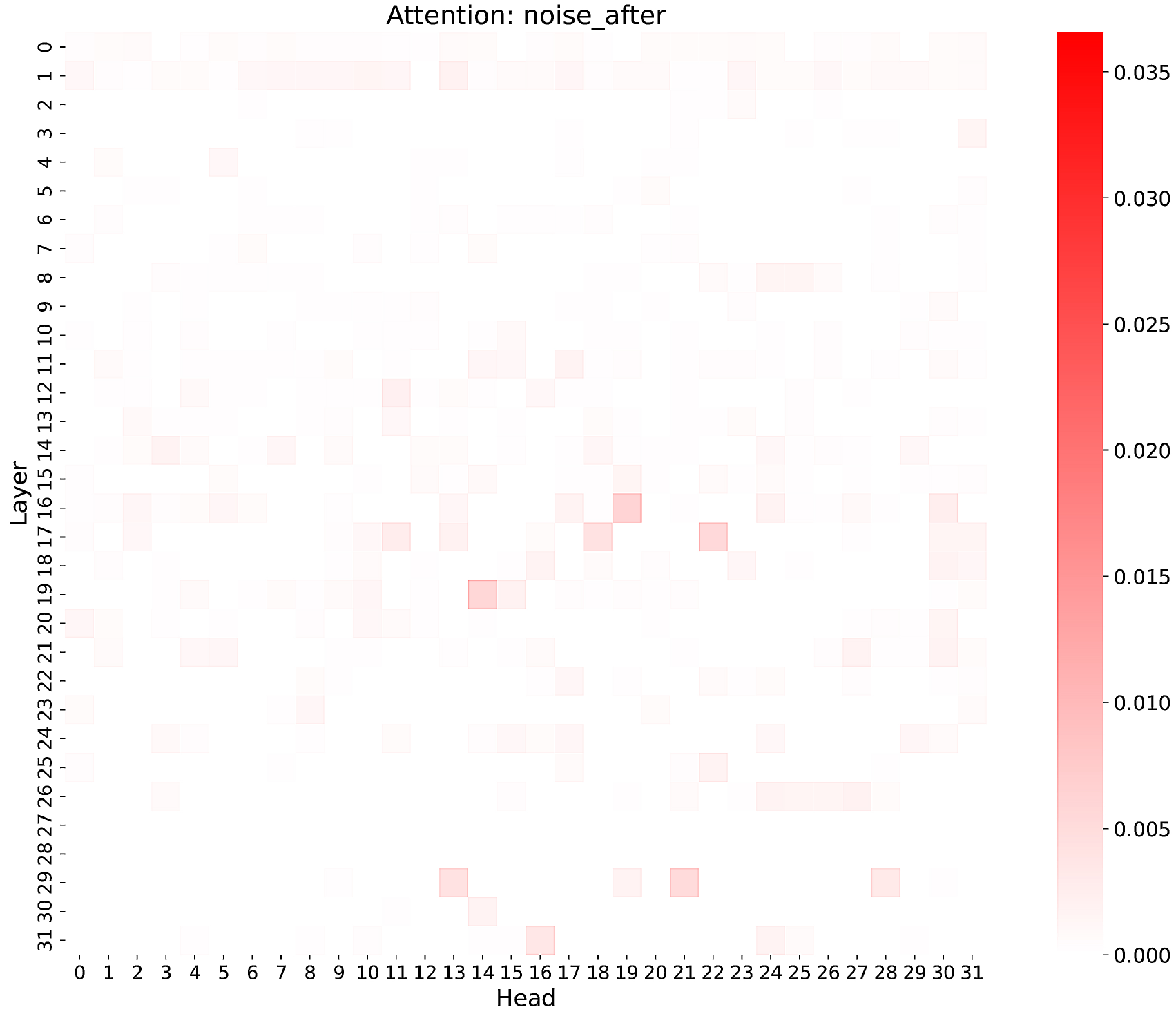}
      \caption{The overall attention heatmaps when inserting noise after relevant documents}
      \label{fig-overallAtten-c}
  \end{subfigure}%
\caption{The overall attention heatmaps across samples, comparing with inserting noise tokens.}
  \label{fig:OverallAttend}
\end{figure}

\subsection{Aggregated Attention Heatmaps Across Examples}
\label{sec:aggregateHeatmaps}

The previous section illustrates attention behaviours on individual examples. To investigate whether these patterns generalize across real-world long documents, we conduct a dataset-level aggregation analysis. Specifically, we sample 500 query–relevant document pairs from the TREC DL19 document ranking's development set (MS MARCO) \cite{{craswell2020overview}}. The RankLLaMA model fine-tuned for document-level reranking\footnote{\url{https://huggingface.co/castorini/rankllama-v1-7b-lora-doc}} is used for evaluation. All documents are truncated to 1200 tokens.

For each example, we compute the average attention scores from document tokens to query tokens across all heads in every layer.
The attention scores are then averaged across the sampled set to produce an overall heatmap summarizing attention behaviour across layers and heads.

To assess robustness against irrelevant context, we also create two additional test settings. For each query–relevant document pair, we randomly sample a negative document, extract its first 800 tokens as noise, and insert the noise either before or after the relevant document. The aggregated attention heatmaps under these two noise injection settings are similarly computed.
The resulting heatmaps for the clean and noisy cases are shown in Fig.~\ref{fig:OverallAttend}. 

\subsubsection{Findings}  
As shown in Fig.~\ref{fig-overallAtten-a}, certain heads—primarily located in the beginning and middle layers, as well as several heads in the final layers—exhibit strong attention from document tokens to query tokens, aligning with relevance-focused behaviour observed earlier. However, when noise is inserted, both Fig.~\ref{fig-overallAtten-b} and Fig.~\ref{fig-overallAtten-c} display a general attenuation of attention scores, indicating that irrelevant content weakens the focus on relevant tokens. Notably, inserting noise \emph{after} the relevant content appears to cause greater attention dispersion, likely because the model processes text autoregressively from left to right.

These aggregated patterns confirm that a fraction of attention heads identify relevance signals, and that irrelevant text can dilute these signals, motivating selective block extraction before LLM reranking.

\subsection{Quantitative Analysis with Attention-Relevance Alignment Score (ARAS) and Positive Correlation Rate (PCR)}
\label{sec:aras_pcr}

To complement the above qualitative attention observations and provide a more systematic and quantitative understanding of how LLM rerankers behave (addressing \textbf{RQ1}), we propose two evaluation metrics: the \textbf{Attention-Relevance Alignment Score (ARAS)} and the \textbf{Positive Correlation Rate (PCR)}. These metrics aim to measure how well the model’s attention aligns with true relevance signals in long documents.

\subsubsection{Metric Definitions}

\begin{itemize}
  \item Attention weight per chunk: Given a query-document pair $(q, d)$, we first segment the document $d$ into $M$ non-overlapping chunks $\{C_1, C_2, \dots, C_M\}$, where each chunk contains a fixed number of tokens. For a given attention head and layer, let $\mathbf{A} \in \mathbb{R}^{L \times L}$ denote the attention matrix for the entire input sequence of length $L$. 

Let $Q = \{q_1, q_2, \dots, q_{|Q|}\}$ represent the token indices of the query portion in the input sequence. Then, for each document chunk $C_i$ (corresponding to token indices $\{c_{i,1}, \dots, c_{i,K}\}$), we compute its average attention weight toward the query tokens as:

\[
\text{AttentionWeight}(C_i) = \frac{1}{K} \sum_{t \in C_i} \frac{1}{|Q|} \sum_{j \in Q} \mathbf{A}_{t, j}
\]

This reflects how much attention the document chunk as a whole assigns to the query tokens.

\item Relevance score per chunk: For each chunk $C_i$, we estimate its relevance with respect to the query using an external cross-encoder model\footnote{\url{https://huggingface.co/cross-encoder/ms-marco-MiniLM-L6-v2}} which serve as an approximate "ground-truth" relevance score:

\begin{multline*}
\text{RelevanceScore}(C_i)\\
= \text{CrossEncoder}(q, \text{text}(C_i))
\end{multline*}

where $\text{text}(C_i)$ denotes the surface text of chunk $C_i$.

\item Attention-Relevance Alignment Score (ARAS): For each query-document pair $(q, d)$, the ARAS is defined as the Spearman rank correlation coefficient between the attention weights and relevance scores across all chunks:
\begin{equation}
\label{eq:aras}
\begin{aligned}
\mathrm{ARAS}(q,d)
&= \mathrm{Spearman}\!\left(\mathbf{a}, \mathbf{r}\right), \\
\mathbf{a} &= \big(\mathrm{Attn}(C_i)\big)_{i=1}^{M}, \\
\mathbf{r} &= \big(\mathrm{Rel}(C_i)\big)_{i=1}^{M}.
\end{aligned}
\end{equation}

This measures how well the ranking induced by attention weights aligns with the ranking of relevance scores across chunks.
For reporting, we compute the overall mean ARAS across the samples:

\[
\overline{\text{ARAS}} = \frac{1}{N} \sum_{j=1}^{N} \text{ARAS}(q_j, d_j)
\]

This average ARAS indicates the typical alignment strength per query-document pair.

\item Positive Correlation Rate (PCR): Given a collection $\mathcal{D} = \{(q_j, d_j)\}_{j=1}^N$ of $N$ query-document pairs, we compute PCR as the proportion of pairs whose ARAS is positive:

\begin{equation}
\label{eq:pcr}
\text{PCR} = \frac{1}{N} \sum_{j=1}^{N} \mathbb{I}\left( \text{ARAS}(q_j, d_j) > 0 \right)
\end{equation}

where $\mathbb{I}(\cdot)$ is the indicator function. PCR reflects how consistently attention correlates positively with relevance across the dataset.

\end{itemize}

ARAS focuses on the alignment strength between attention and relevance for individual query-document pairs, while PCR reflects alignment stability across the dataset. Together, these metrics offer both instance-level and corpus-level interpretability.

\subsubsection{Experimental Setup}

\begin{figure*}[!p]
  \centering
  \begin{subfigure}[t]{0.43\linewidth}
      \centering
      \includegraphics[width=\linewidth]{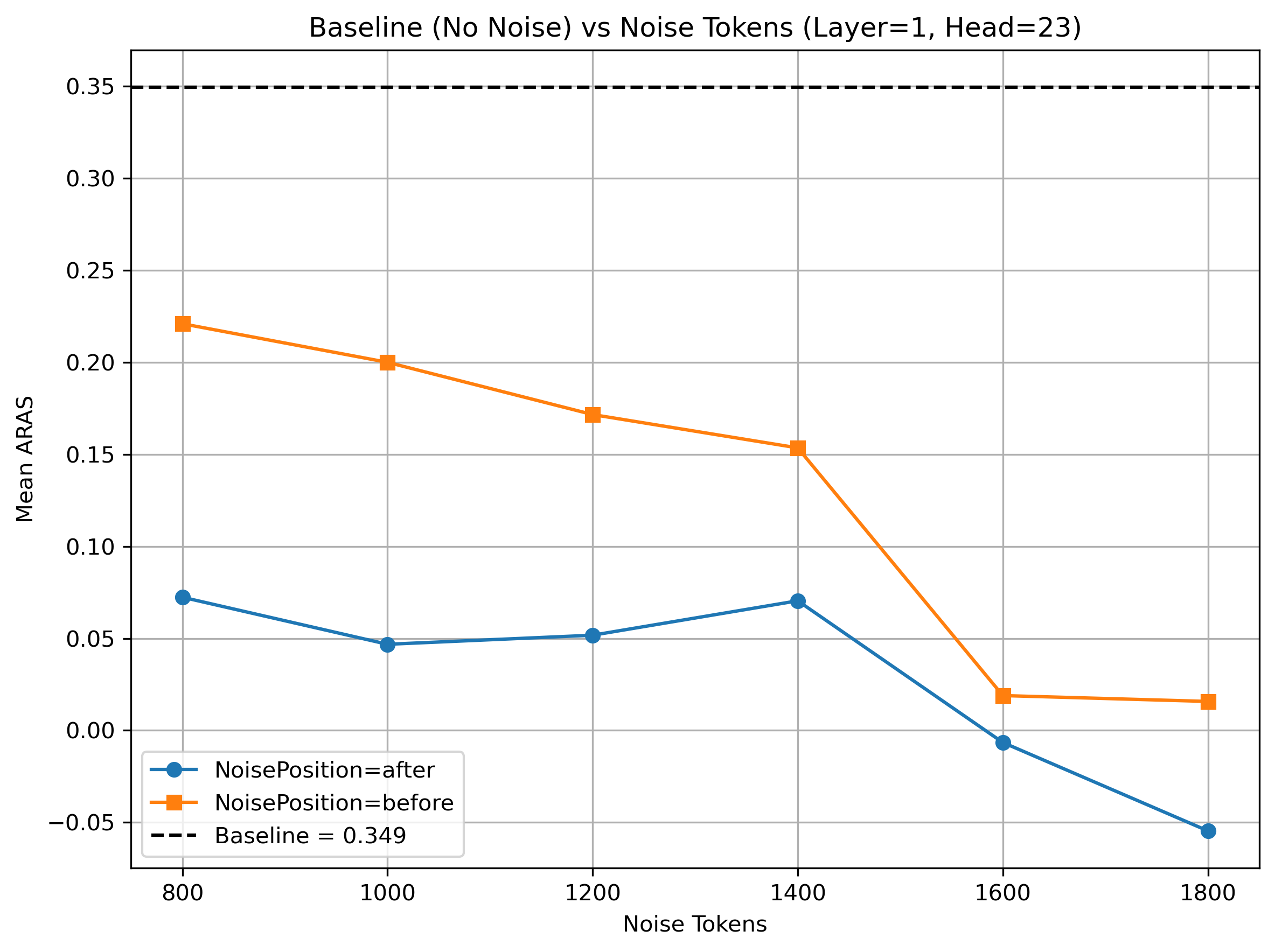}
      \caption{ARAS of Layer 1 Head 23}
      \label{fig:layer8ARAS-1}
  \end{subfigure}
  \hspace{25pt}
  \begin{subfigure}[t]{0.43\linewidth}
      \centering
      \includegraphics[width=\linewidth]{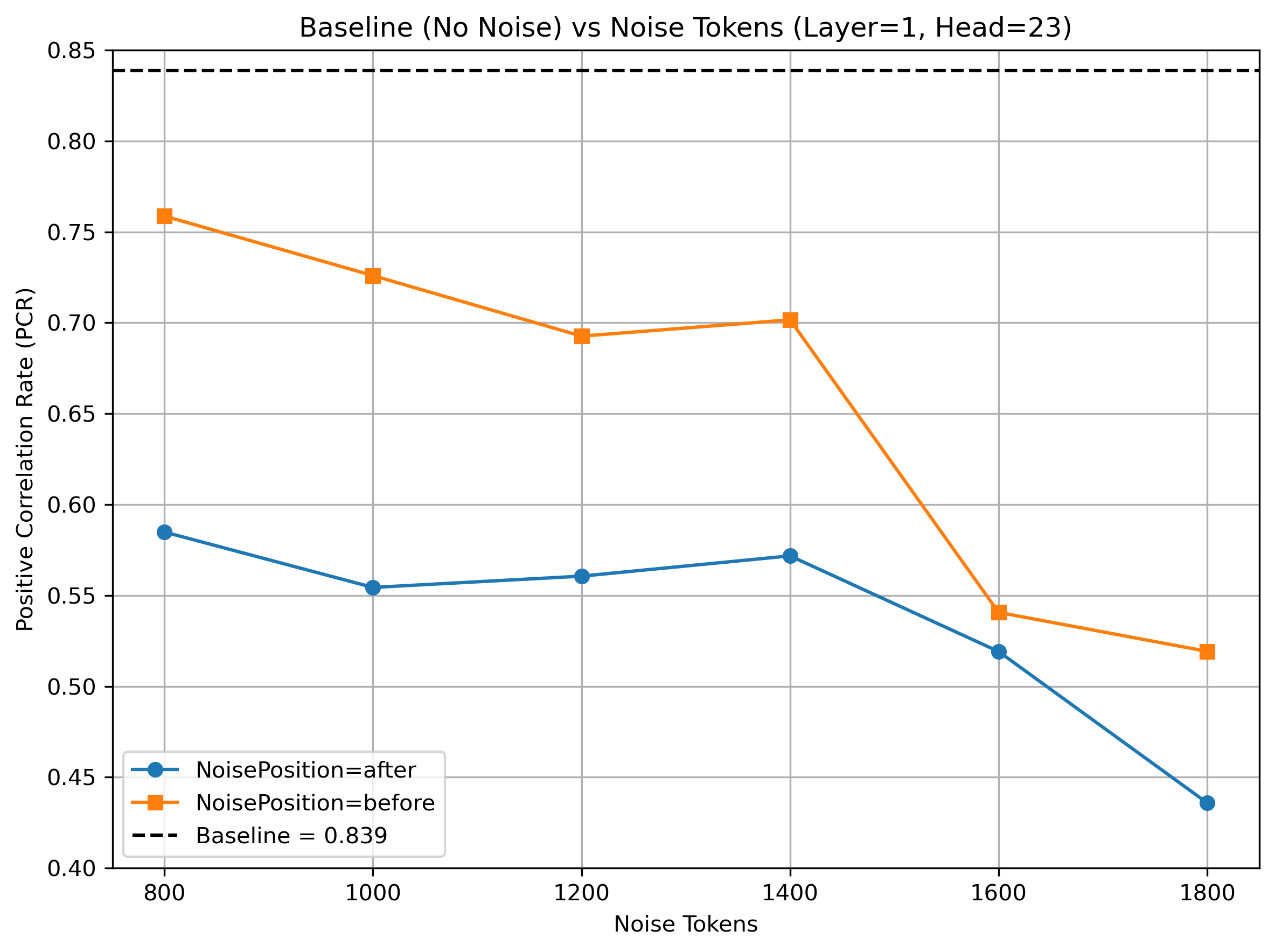}
      \caption{PCR of Layer 1 Head 23}
      \label{fig:layer8PCA-1}
  \end{subfigure}

  \centering
  \begin{subfigure}[t]{0.43\linewidth}
      \centering
      \includegraphics[width=\linewidth]{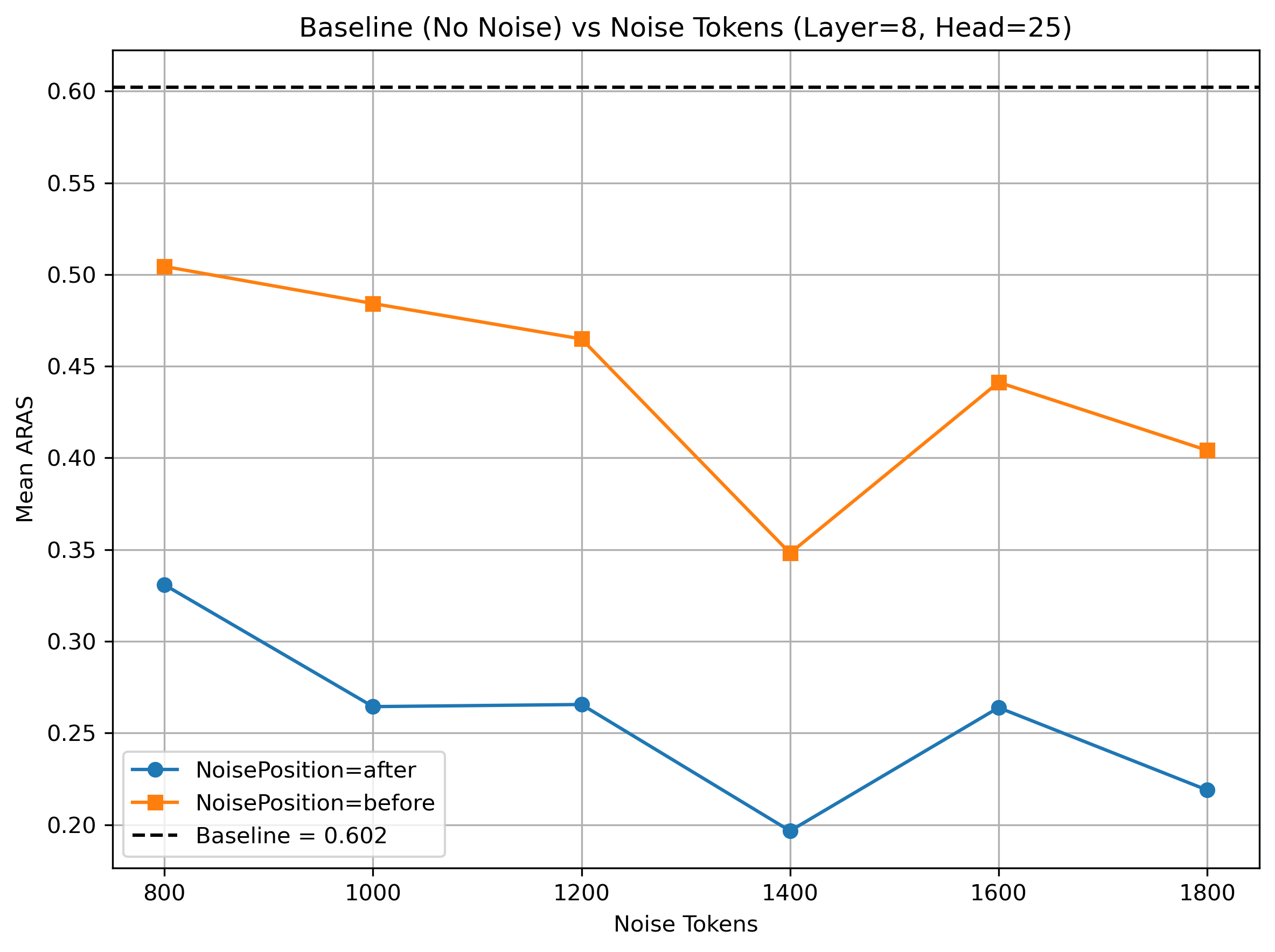}
      \caption{ARAS of Layer 8 Head 25}
      \label{fig:layer8ARAS-2}
  \end{subfigure}
  \hspace{25pt}
  \begin{subfigure}[t]{0.43\linewidth}
      \centering
      \includegraphics[width=\linewidth]{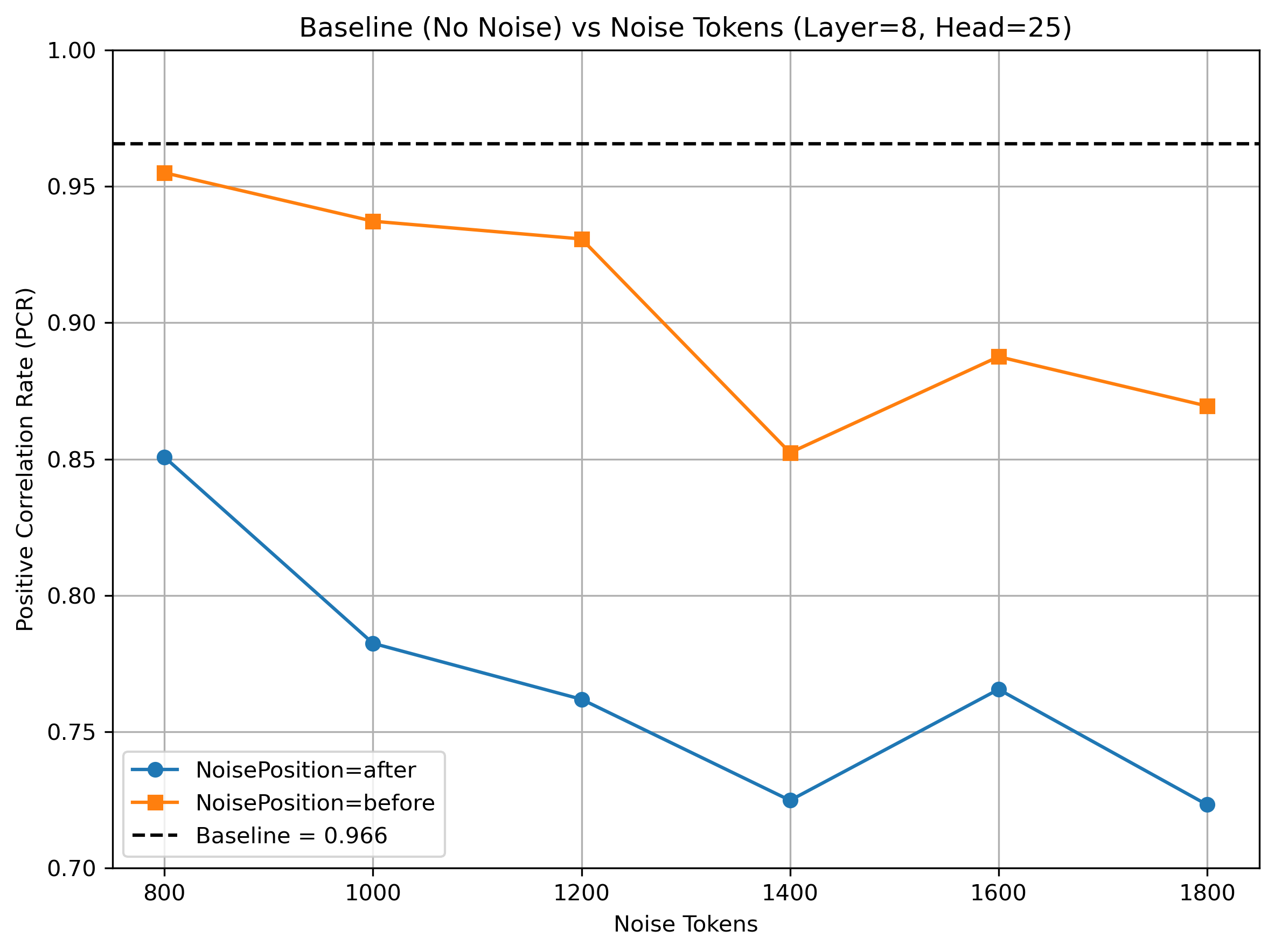}
      \caption{PCR of Layer 8 Head 25}
      \label{fig:layer8PCA-2}
  \end{subfigure}

  \centering
  \begin{subfigure}[t]{0.43\linewidth}
      \centering
      \includegraphics[width=\linewidth]{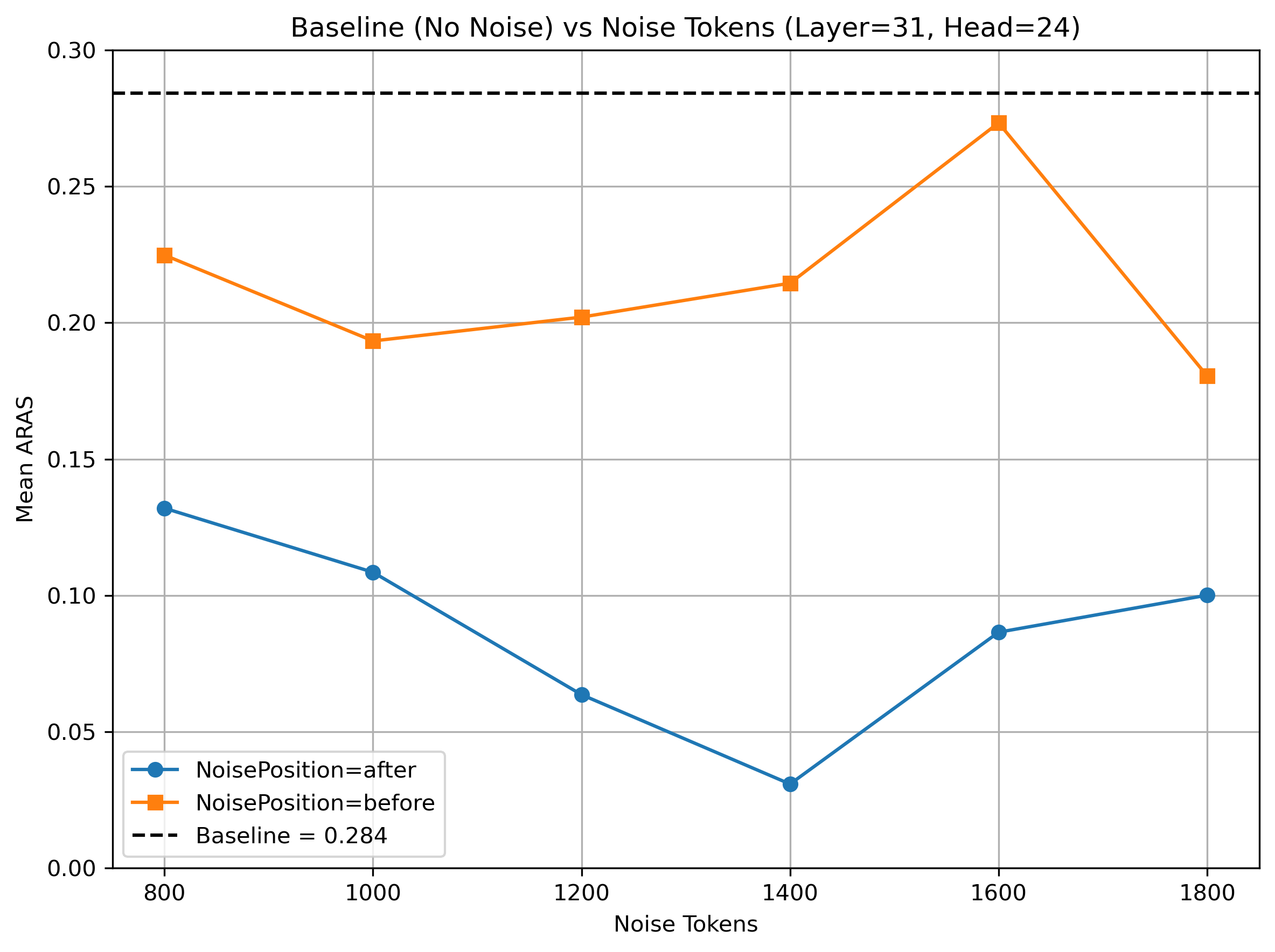}
      \caption{ARAS of Layer 31 Head 24}
      \label{fig:layer8ARAS-3}
  \end{subfigure}
  \hspace{25pt}
  \begin{subfigure}[t]{0.43\linewidth}
      \centering
      \includegraphics[width=\linewidth]{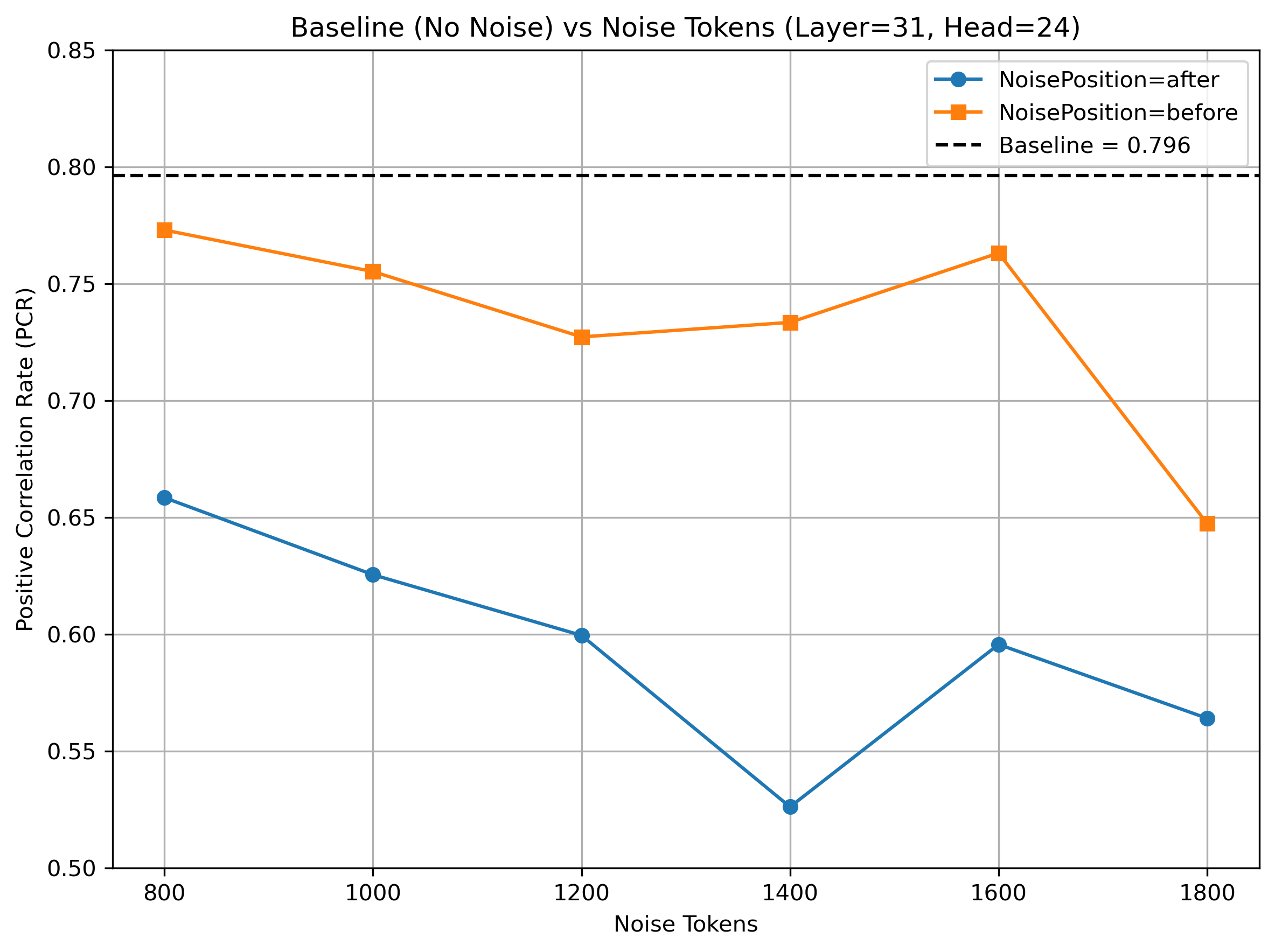}
      \caption{PCR of Layer 31 Head 24}
      \label{fig:layer8PCA-3}
  \end{subfigure}

  \caption{Mean ARAS scores and positive correlation rates  for different layers across various noise lengths and different noise insertion positions. Baseline is the score without inserting noise tokens.}
  \label{fig:ARAS_PCR}
\end{figure*}

We conduct controlled experiments on 500 query-relevant document pairs sampled from the same development set (MS MARCO) as Section \ref{sec:aggregateHeatmaps}.
Each document is truncated to 1200 tokens, and segmented into 64-token chunks for analysis. The ARAS and PCR metrics are then computed across three representative attention heads previously identified: Layer 1 Head 23, Layer 8 Head 25, and Layer 31 Head 24. The figures are shown in Fig.~\ref{fig:ARAS_PCR}.

To simulate long irrelevant contexts, we insert 800 to 1800 noise tokens either before or after the relevant document content, allowing us to test attention stability under varying noise levels.

\subsubsection{Results and Findings}

\paragraph{Layer 1, Head 23.}
Under clean input, ARAS achieves 0.349 and PCR reaches 83.9\%, suggesting moderate attention-relevance alignment at shallow layers. However, performance degrades significantly as noise is inserted, particularly when noise follows the relevant content. ARAS drops to negative values and PCR falls below 50\% under heavy noise, indicating this head is highly sensitive to positional disruptions.

\paragraph{Layer 8, Head 25.}
This head shows the strongest relevance alignment: baseline ARAS reaches 0.602 and PCR 96.6\%. Although both metrics decline with added noise, PCR remains relatively stable (above 70\%), suggesting that this mid-layer head is better at resisting irrelevant information. Nevertheless, ARAS still declines sharply as more noise accumulates, 
indicating that although the head consistently identifies relevant regions, the precision of its attention distribution becomes less aligned with true relevance as noise increases.

\paragraph{Layer 31, Head 24.}
This deep-layer head exhibits relatively weak alignment, with a baseline ARAS of 0.284 and PCR of 79.6\%. As with earlier heads, both metrics decline in the presence of noise. Although its performance degrades more slowly than Layer 1, it still suffers substantial drops under high noise levels: ARAS falls below 0.1 and PCR drops to approximately 56\%.

\paragraph{Overall Observations.}
Across all heads, inserting irrelevant tokens \textit{after} the relevant content consistently causes more severe alignment degradation than inserting noise \textit{before}, likely because autoregressive LLMs encounter relevant content later, reducing the usable attention capacity. 

\subsection{Summary: Implications for Long Document Retrieval}

These findings provide a direct answer to \textbf{RQ1}, showing that while certain attention heads do exhibit relevance-focused behaviors, their ability to preserve this alignment diminishes substantially when irrelevant content accumulates. This degradation is especially pronounced when noise appears later in the sequence, likely due to the left-to-right processing nature of decoder-only LLMs such as RankLLaMA.

These results reinforce the need for explicit block selection before LLM reranking—not only to reduce computational overhead, but also to preserve attention focus, mitigate distraction from irrelevant text, and prevent attention dispersion. This insight validates the core motivation behind EviRerank: block selection remains necessary even in the era of large language models. 
By selecting and ranking the most relevant content blocks, we enable LLMs to concentrate attention on relevant information, thereby improving both retrieval effectiveness and computational efficiency.

\end{document}